\documentclass[twocolumn,tighten]{aastex62}

\newcommand{\msun}{M$_\odot$}

\begin{document}

\title{On the Absence of High-Redshift AGNs: Little Growth in the Supermassive 
Black Hole Population at High Redshifts}

\author[0000-0002-6319-1575]{L.~L.~Cowie}
\affiliation{Institute for Astronomy, University of Hawaii,
2680 Woodlawn Drive, Honolulu, HI 96822, USA}

\author[0000-0002-3306-1606]{A.~J.~Barger}
\affiliation{Department of Astronomy, University of Wisconsin-Madison,
475 N. Charter Street, Madison, WI 53706, USA}
\affiliation{Department of Physics and Astronomy, University of Hawaii,
2505 Correa Road, Honolulu, HI 96822, USA}
\affiliation{Institute for Astronomy, University of Hawaii, 2680 Woodlawn Drive,
Honolulu, HI 96822, USA}

\author{F.~E.~Bauer}
\affiliation{Instituto de Astrof\'isica and Centro de Astroingenier\'ia, Facultad de F\'isica, 
Pontificia Universidad Cat\'olica de Chile,
Casilla 306, Santiago 22, Chile}
\affiliation{Millennium Institute of Astrophysics (MAS), Nuncio Monse{\~{n}}or S{\'{o}}tero 
Sanz 100, Providencia, Santiago, Chile} 
\affiliation{Space Science Institute,
4750 Walnut Street, Suite 205, Boulder, Colorado 80301, USA} 

\author[0000-0003-3926-1411]{J.~Gonz{\'a}lez-L{\'o}pez}
\affiliation{N\'ucleo de Astronom\'ia de la Facultad de Ingenir\'ia y Ciencias, Universidad Diego Portales,
Av. Ej\'ercito Libertador 441, Santiago, Chile}
\affiliation{Instituto de Astrof\'isica and Centro de Astroingenier\'ia, Facultad de F\'isica, 
Pontificia Universidad Cat\'olica de Chile \\
Casilla 306, Santiago 22, Chile}

\begin{abstract}
We search for high-redshift ($z>4.5$) X-ray AGNs
in the deep central (off-axis angle $<5\farcm7$) region of the 7~Ms
{\em Chandra\/} Deep Field-South X-ray image. We compile an initial
candidate sample from direct X-ray detections.  We then probe
more deeply in the X-ray data by
using pre-selected samples with high spatial resolution NIR/MIR
({\em HST\/} 1.6~$\mu$m and {\em Spitzer\/} 4.5~$\mu$m) and submillimeter
(ALMA 850~$\mu$m) observations. The combination of the NIR/MIR
and submillimeter pre-selections
allows us to find X-ray sources with a wide range of dust properties
and spectral energy distributions (SEDs). We use the SEDs
from the optical to the submillimeter to determine if previous photometric redshifts
were plausible. Only five possible $z>5$ X-ray AGNs  are found, all of which
might also lie at lower redshifts. If they do lie at high redshifts, then two are
Compton-thick AGNs, and three are ALMA 850~$\mu$m sources.
We find that (i) the number density of X-ray AGNs is dropping rapidly at high
redshifts, (ii) the detected AGNs do not contribute significantly
to the photoionization at $z>5$, and (iii) the measured X-ray light density over $z=5-10$
implies a very low black hole accretion density with very little growth
in the black hole mass density in this redshift range.
\end{abstract}


\section{Introduction}
\label{secintro}
\begin{figure}[ht]
\centerline{\includegraphics[width=9cm,angle=0]{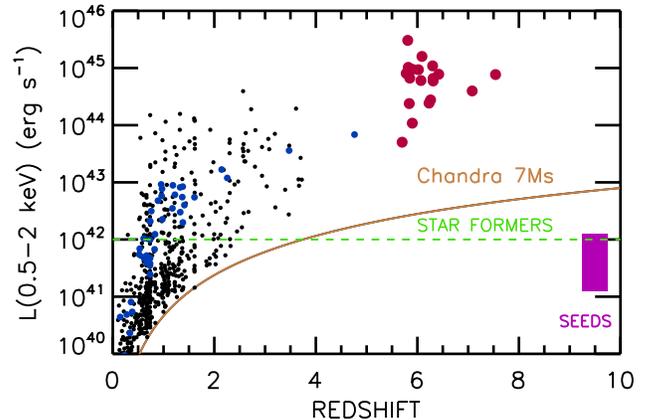}}
\caption{$L_{0.5-2~{\rm keV}}$ (see Equation~1) vs. redshift for
spectroscopically identified sources in the CDF-S (black circles). 
The gold curve shows the limit corresponding to 
$f_{0.5-2~{\rm keV}}=10^{-17}$~erg~cm$^{-2}$~s$^{-1}$. 
When a source is undetected in 
0.5--2~keV, we plot $L_{2-7~{\rm keV}}$ (see Equation~2) in blue. 
We also show the rough luminosity (green dashed line) below which 
X-ray contributions from star formation may become significant.
The purple shaded region shows the possible luminosities of 
$10^4$--$10^5$~\msun\ direct
collapse seeds, if these radiate at or near the Eddington limit.
We assume 10\% of the bolometric luminosity is in the
observed-frame 0.5--2~{\rm keV} band (Ricarte et al.\ 2019).
The red large circles show X-ray detected sources above $z=5.5$
(Nanni et al.\ 2017; Ba\~nados et al.\ 2018)
that were selected from very wide-area optical surveys.
\label{5ms_lums}
}
\end{figure}

While large samples of $z\sim1$--5 active galactic nuclei (AGNs) 
have now been assembled (e.g., Brandt \& Alexander 2015),
our information on $z>5$ AGNs remains far more limited. 
Only about 30 sources have been found beyond $z=6.5$ (Pons et al. 2019), and
only two beyond $z=7$ (Ba\~nados et al.\ 2018).
While some hundreds of AGNs have been spectroscopically identified
beyond $z=5.5$ in the rest-frame optical (e.g., Ba\~nados et al.\ 2016), 
and some tens of these subsequently observed in X-rays (e.g.,
Nanni et al.\ 2017; Vito et al.\ 2019a, 2019b; Salvestrini et al.\ 2019), 
these primarily arise from the extremely luminous tail of the population
(see Figure~\ref{5ms_lums}). Direct X-ray searches in the ultradeep
{\em Chandra\/} Deep Fields (CDFs) can detect much fainter
AGNs, but they have yielded only one spectroscopically
identified source beyond $z=5$: the $z=5.186$ source found by
Barger et al.\ (2002, 2003) in the CDF-N. The CDF-S 7~Ms
X-ray image (Luo et al.\ 2017; hereafter, L17) is sensitive enough in its central
regions to detect AGNs with observed-frame 0.5--2~keV luminosities
$\gtrsim10^{42.5}$~erg~s$^{-1}$ through much of the $z=5$--10 range
(see Figure~\ref{5ms_lums}), but here, again, there are few spectroscopically 
identified high-redshift sources, with only one at 
$z>4.5$: the Compton-thick $z=4.762$ source found by Gilli et al.\ (2011, 2014).
In the larger but shallower COSMOS field, there are only two sources with 
spectroscopic redshifts beyond $z=5$ (Marchesi et al.\ 2016), with the highest redshift
source being the $z=5.3$ source of Capak et al.\ (2011). 

The central regions of both CDFs lie in the CANDELS/GOODS 
areas where the deep {\em HST\/} (Giavalisco et al.\ 2004)
and {\em Spitzer\/} (Dickinson et al.\ 2003; Ashby et al.\ 2013; 
Labb{\'e} et al.\ 2015)
data allow good photometric 
redshifts (hereafter, photzs) to be derived. However, even photzs 
yield a very small number of potential high-redshift AGNs 
(e.g., Giallongo et al.\ 2015, hereafter, G15; Cappelluti et al.\ 2016;
Weigel et al.\ 2015; Pacucci et al.\ 2016; Parsa et al.\ 2018,
 Giallongo et al.\ 2019), and, 
as we shall discuss, at least some of these are simply
photz errors that have placed the sources at too high of redshifts.

Despite these observational results, the $z=5$--10 interval is widely considered to be 
a key period in the growth of supermassive black holes (SMBHs) when accretion
and merging turn seed black holes at $z > 10$ into the SMBHs seen at later 
redshifts. This process is most dramatically constrained by the sources hosting
$\sim10^9$~\msun\ SMBHs at $z=6$--7;
their seeds are theorized to be either $\lesssim100$~\msun\ Population~III
stellar remnants (e.g., Madau \& Rees 2001; Volonteri, Haardt \& Madau 2003) or 
$\sim10^5$~\msun\ Direct Collapse Black Holes (DCBHs)
formed from the collapse of primordial gas clouds
(e.g., Haehnelt \& Rees 1993; Bromm \& Loeb 2003; Lodato \& Natarajan 2006).

In this paper, we aim to assess whether current X-ray data on the CDF-S, 
together with optical through far-infrared (FIR)/submillimeter data, imply that there is very 
little growth in the overall SMBH population (as opposed to in a small number of 
extremely luminous sources) during this period. We will argue that there are too few 
detected high-redshift AGNs to account for any major SMBH growth,
and these sources do not contribute significantly
to the photoionization at these redshifts.

We use $(\Omega_m, \Omega_\Lambda, h) = (0.32, 0.68, 0.67)$ from the 
Planck Collaboration VI (2018).

\section{Redshifts and Luminosities for Direct X-ray Detections}
\label{secdir}
In Barger et al.\ (2019), we presented a critically reviewed 
spectroscopic catalog of the 7~Ms X-ray observations of the CDF-S that was 
based on the L17 X-ray sample restricted to off-axis angles $<10'$,  
giving a sample of 938 X-ray sources. Only the L17 directly X-ray detected 
sources are included, and not the L17 supplemental catalog, which uses 
near-infrared (NIR) pre-selection. We consider the X-ray properties of a 
NIR/mid-infrared (MIR) pre-selected sample in Section~\ref{nir_select}.

L17 determined the most probable optical/NIR counterparts to the 
X-ray sources based on both the $2\sigma$ positional uncertainties 
of the X-ray sources and the magnitudes of the potential counterparts.
We adopt these identifications. The spectroscopic identifications are 
heavily overlapped with previous summaries, such as those of L17 and 
Vito et al.\ (2018), but differ for a number of sources.
In Barger et al.\ (2019), we also provided spectroscopic classifications for the X-ray 
sources based on their optical/NIR spectra (see their Table~1).
In total, we have secure redshifts (hereafter, speczs) for 605 of the 938 
sources, or 64.5\%, and spectroscopic classifications for 596 of these. 

We computed the observed-frame soft (0.5--2~keV) and hard (2--7~keV) 
luminosities from
\begin{equation}
L_{0.5-2~{\rm keV}} = 4\pi d_L^2 f_{0.5-2\,{\rm keV}} (1+z)^{\Gamma-2}~{\rm erg~s^{-1}} \,,
\end{equation}
and
\begin{equation}
L_{2-7\,{\rm keV}} = 4\pi d_L^2 f_{2-7\,{\rm keV}} (1+z)^{\Gamma-2}~{\rm erg~s^{-1}} \,,
\end{equation}
where $d_L$ is the luminosity distance, and $f_{0.5-2\,{\rm keV}}$ or
$f_{2-7\,{\rm keV}}$ is the observed flux in that band.
We take the  photon index to be $\Gamma=1.8$ for all.
These luminosities do not account for absorption, but at high redshifts, they correspond
to very high rest-frame energies (at $z=5$, 3--12~keV and 12--42~keV, respectively,
for the two bands), where absorption effects should be minimal, except in the case of the 
most extreme sources (i.e., Compton thick or near-Compton thick). 
We consider this further in the discussion.

We show the distribution of luminosity versus redshift for the sources with
speczs in Figure~\ref{5ms_lums}, 
where the strong cut-off above $z\sim4$ can be clearly seen, despite 
the fact that luminous X-ray AGNs could easily be detected at these redshifts.

For the present analysis, we restrict our sample
even further to off-axis angles $<5\farcm7$, thereby selecting the deepest
part of the {\em Chandra\/} image and also where the {\em Chandra\/}
point spread function (PSF) is reasonably uniform. As we shall
discuss in Section~\ref{alma_select}, this is also
the region where there are extremely deep {\em Herschel\/} 
and ground-based submillimeter observations.
There are 526 X-ray sources in this central region.
In the soft band (0.5--2~keV), the sensitivity rises from an on-axis value
of $7\times10^{-18}$~erg~cm$^{-2}$~s$^{-1}$ to 
$2\times10^{-17}$~erg~cm$^{-2}$~s$^{-1}$ at $5\farcm7$
and to just over $10^{-16}$~erg~cm$^{-2}$~s$^{-1}$
at the $10'$ outer radius (L17). In the hard band (2--7~keV), the 
sensitivities are about a factor of 5 higher.

Because of the extensive high-quality optical/NIR/MIR data
in the CDF-S, numerous groups have estimated photzs in the region
(e.g., Santini et al.\ 2009, 2015; Rafferty et al.\ 2011;
Dahlen et al.\ 2013; Hsu et al.\ 2014; Skelton et al.\ 2014; Straatman et al.\ 2016).
We consider here the most recent of these catalogs, and, in particular,
the Hsu et al.\ (2014; hereafter, H14), 
Santini et al.\ (2015\footnote{Santini et al.\ (2015) published the official 
CANDELS photzs derived using techniques described in Dahlen et al.\ (2013).}; 
hereafter, CANDELS), and Straatman et al.\ (2016; hereafter, ZFOURGE)
results, all of which use the deep {\em Spitzer\/} IRAC data of 
Ashby et al.\ (2013) or Labb{\'e} et al.\ (2015).

Vito et al.\ (2018) concluded that
the ZFOURGE photzs provided the best approximation to the speczs
in the L17 sample, while Barger et al.\ (2019) favored the H14 photzs, which 
are based on galaxy/AGN templates and give the fewest outliers.
Vito et al.\ (2018) used the probability distributions of the photzs 
from ZFOURGE to analyze the redshift 
distributions in the X-ray sample. However, there is considerable scatter in the 
estimates from the different catalogs (at least some of which must arise from
the differences in the adopted templates), so we consider instead the range
in redshifts from the three catalogs as giving an estimate of the
uncertainties. Note that none of the photz codes include
ultraluminous infrared galaxy (ULIRG) templates, which are likely to be the best match 
to heavily obscured AGNs at $z>2$. Thus, when we turn to the FIR/submillimeter 
detected sources in Section~\ref{alma_select}, we will also incorporate redshifts
estimated from FIR spectral energy distributions (SEDs) to test the photz estimates.

Adopting the photzs from H14, when available, and otherwise those from ZFOURGE
(note that we do not apply any of their quality restrictions to their photzs),
all but 12 of the X-ray sources in the central region have either speczs or photzs.
These 12 sources do not appear in the optical/NIR catalogs
and thus also do not have CANDELS photzs. In L17, eight of these are described
as unmatched sources and four as matched sources to CANDELS. However, 
none of the four claimed matched sources are present in the Guo et al.\ (2013) catalog.
Moreover, none of the 12 sources are visible in either the 1.6~$\mu$m image or 
the ultradeep 4.5~$\mu$m image of Labb{\'e} et al.\ (2015).

Many of these blank or very faint optical/NIR sources may be false positives in L17.
For example, Vito et al.\ (2018) used false probabilities greater or approximately
equal to $10^{-4}$ to separate false positives. If we adopt this
criterion, then 9 of the 12 are false positives.
These 9 sources are likely not real X-ray detections, and
we do not consider them further in this paper.
However, the remaining 3 sources have lower false probabilities
(L17~\#226, \#238, \#492).
One of these sources (L17~\#492) lies within $1\farcs5$
of a bright galaxy, which the CANDELS photz places at low redshift.
We consider it likely that the X-ray source is associated with this bright galaxy.
However, the remaining two sources are blank from the optical
through the 24~$\mu$m and also have no submillimeter counterpart.
While L17~\#226 has a logarithmic false probability of
$-4.3$ and could be spurious, L17~\#238 has a logarithmic false probability
of only $-6.4$ and appears real in the X-ray images.
L17~\#238 is also the only one of the 12
with a radio counterpart in the Miller et al.\ (2013) catalog.
We are unable to do anything further with these two sources.

Eight of the L17 X-ray catalog sources in the central region are placed 
at $z>4.5$ by at least one of the photz estimates.
These eight are listed in Table~\ref{tab1}, denoted by their L17 
catalog numbers.

\section{Searching for High-Redshift X-ray AGNs Using Pre-Selected Samples}
\label{other_select}
We can potentially probe more deeply into the X-ray population by pre-selecting
samples at other wavelengths. This has most often been done by
choosing samples in the optical/NIR, which allows one to search
for high-redshift AGNs based on colors or photzs. We can also
stack at the source positions of an optical/NIR sample, or 
average the X-ray fluxes or luminosities measured at those positions
to determine the means as a function of redshift. 
Such analyses by Willott (2011),
Cowie et al.\ (2012), and Vito et al.\ (2016) based on the CDF-S 
have failed to detect any significant X-ray signal in the optical/NIR
samples at $z>5$, a result we confirm in Section~\ref{nir_select}.

One possible problem is that the X-ray sources may be 
much redder than the average optical/NIR galaxy 
at these redshifts, and, thus, the highest redshift X-ray AGNs 
may become too faint to detect in the observed-frame optical/NIR. 
We investigate this possibility in Section~\ref{alma_select} by starting
with an ALMA sample. While the sample is small, we do find several
submillimeter sources that are potential high-redshift luminous X-ray AGNs.

\subsection{Measuring X-ray Fluxes at the Positions of Pre-Selected Samples}
\label{xray_preselect}
We measured X-ray fluxes for sources in each of the pre-selected samples
(see  Sections~\ref{nir_select} and \ref{alma_select})
that lie in the central $5\farcm7$ region, where the {\em Chandra\/} data 
are most sensitive, using the X-ray catalog and images provided by L17. 
When a source in the L17 catalog has a counterpart within $1''$, 
we take the X-ray flux from that catalog. 
The matching is relatively insensitive to the choice of matching
radius, which is based on the 2$\sigma$ positional uncertainties 
from L17 of the fainter {\em Chandra\/} sources.

For the remaining sources, we measured both the 
0.5--2~keV and 2--7~keV fluxes following 
the procedure outlined in Cowie et al.\ (2012). 
We used a circular aperture, which provides a good approximation 
to the PSF shape at these small off-axis angles.
We adopted a $1\farcs25$ aperture radius, 
which provides a good compromise between including most of
the counts, maximizing the signal-to-noise (S/N), and minimizing 
the contamination from neighboring sources.
With the aperture specified, we computed the X-ray
counts~s$^{-1}$ from $C = (S-B)/(t$),
where $S$ is the number of counts in the aperture,
$B(=\pi r^2 b)$ is the number of background counts expected
in the same aperture, and $t$ is the effective exposure time at the
position of this aperture. We measured the mean background 
$b$ (counts~arcsec$^{-2}$) in an 8$''$--22$''$ annulus around the 
source after clipping pixels with more than 4 counts. 
(See Cowie et al.\ 2012 for an extensive discussion of this choice.) 
$C$ may be negative or positive.
We converted the counts to fluxes using a single normalization,
which we chose by comparing our aperture fluxes measured
for sources in the L17 catalog with the L17 fluxes.
We found good agreement with a scatter
of 23\%, which is adequate for the present work.
A more complete description, including the error
estimation, may be found in Barger et al.\ (2019).

If a pre-selected source is detected at $>3\sigma$ in either the
0.5--2~keV or 2--7~keV band, then we consider the source 
to be X-ray detected.

\subsection{NIR/MIR Pre-Selection}
\label{nir_select}
We begin with the 1.6~$\mu$m CANDELS catalog of Guo et al.\ (2013).
We only include sources detected above $10\sigma$ in either 
the {\em HST\/} F160W ($1.6~\mu$m) band
or the {\em Spitzer\/} channel~2 ($4.5~\mu$m) to optimize the 
quality of the photzs, and
we restrict to the central $5\farcm7$ region.
We hereafter refer to this as our NIR sample.
We measured the X-ray fluxes at the positions of this sample 
as described in Section~\ref{xray_preselect}. 
Notably, none of the 39 galaxies in the central region with 
$z_{\rm spec}>4.5$ is X-ray detected. 

\begin{figure}[ht]
\centerline{\includegraphics[width=9cm,angle=0]{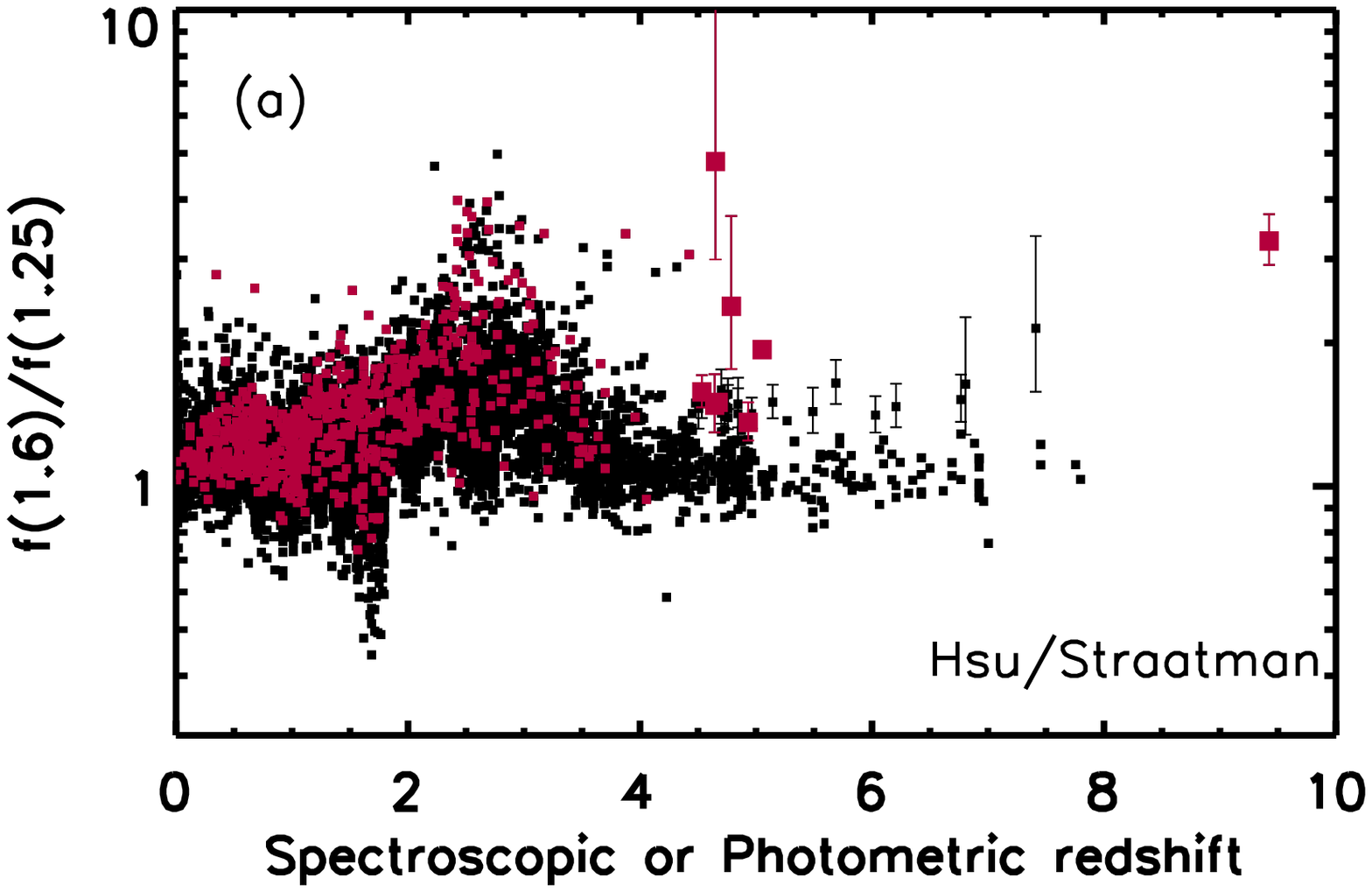}}
\centerline{\includegraphics[width=9cm,angle=0]{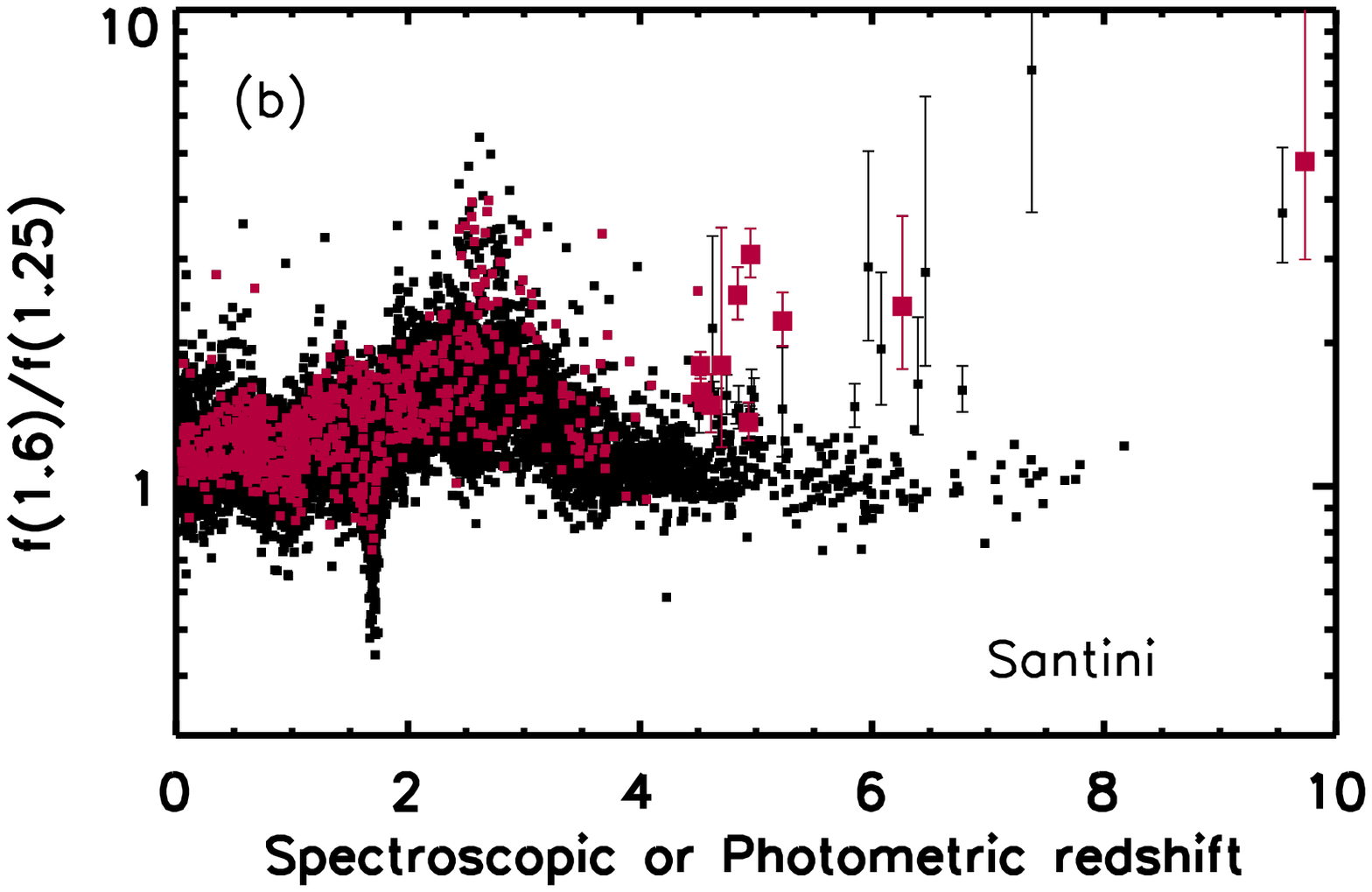}}
\caption{(a) $1.6~\mu$m to $1.25~\mu$m flux ratio vs. specz or
photz for the sources in the Guo et al.\ (2013) $1.6~\mu$m CANDELS 
catalog that lie in the central region and are detected above
$10\sigma$ in either {\em HST\/} F160W ($1.6~\mu$m) or 
{\em Spitzer\/} channel~2 ($4.5~\mu$m) (black squares). 
In (a), we use the H14 photzs, when available, then 
the ZFOURGE photzs (Straatman et al.\ 2016), when available. 
In (b), we use the CANDELS photzs (Santini et al.\ 2015).
The red squares show X-ray detected sources,
with larger squares denoting those at $z_{\rm phot}>4.5$.
For clarity, we only show $1\sigma$ error bars for the redder sources 
at $z_{\rm phot}>4.5$. 
\label{sant_fs} 
}
\end{figure}

In Figure~\ref{sant_fs}, we plot the F160W over F125W flux ratio
versus redshift (specz, when available, or otherwise photz) for our NIR sample.
In (a), we show the photzs from H14 and ZFOURGE, and in (b), 
we show the photzs from CANDELS. 
We show X-ray detected sources in red and the remaining sources in black.

Above $z=4$, most sources have flux ratios near 1, reflecting 
the flat $f_\nu$ of unobscured star-forming galaxies in the near-UV. 
At lower redshifts, there is a wide spread in colors, reflecting the wide range 
of galaxy types. Importantly, all the X-ray detected $z_{\rm phot}$ 
candidates in Figure~\ref{sant_fs} (red large squares) 
are extremely red in the NIR bands.
In Figure~\ref{sant_fs}(a), there are eight X-ray detected 
$z_{\rm phot}>4.5$ candidates,
while in (b), there are 11. In (a), only 2 lie above $z=5$ and 1 above $z=6$,
while in (b), 3 lie above $z=5$ and 2 above $z=6$.
Pacucci et al.\ (2016) used the CANDELS photzs 
to search for high-redshift AGNs, and they considered these 
two $z_{\rm phot}>6$ sources to be candidate DCBHs. 
We discuss them in more detail below.  

\begin{figure}[ht]
\centerline{\includegraphics[width=9cm,angle=0]{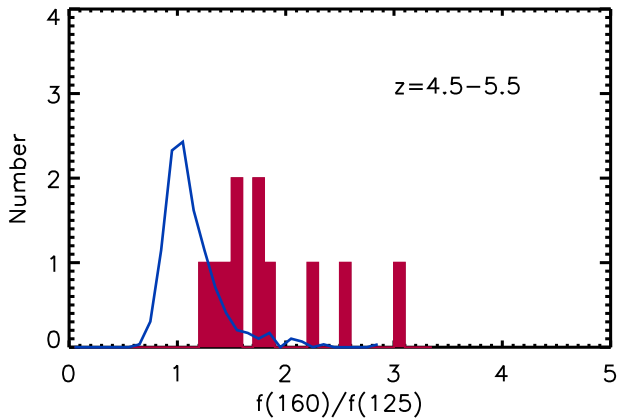}}
\caption{Distribution of the $1.6~\mu$m to $1.25~\mu$m flux
ratio for the sources in the Guo et al.\ (2013) $1.6~\mu$m CANDELS
catalog that lie in the central region, are detected above $10\sigma$
in either {\em HST\/} F160W (1.6~$\mu$m) or {\em Spitzer\/}
channel~2 (4.5~$\mu$m), have $>3\sigma$ detections at 
0.5--2~keV or 2--7~keV, and redshifts in the range $z=4.5$--5.5 
(red histogram) based on the CANDELS photzs (Santini et al.\ 2015).
The blue curve shows the distribution for the non-X-ray sources
in the same redshift range
normalized to have the same number of sources.
\label{color_hist}
}
\end{figure}

If the 11 CANDELS photzs are genuine high-redshift X-ray sources,
then we can see from Figure~\ref{sant_fs}(b) that they are systematically
redder than the non-X-ray population at the same redshift, and, conversely,
that a large fraction of the high-redshift red sources are X-ray sources. 
Based on theoretical modeling, Pacucci et al.\ (2016) argue 
that such red colors may be a good way of finding high-redshift AGNs. 
We can see the red colors of these candidates
more clearly in Figure~\ref{color_hist}, where we compare a 
histogram of the F160W to F125W flux ratios of the X-ray sources 
in the redshift range $z=4.5$--5.5 (the last redshifts where there are
a significant number of sources) (red) to the distribution of non-X-ray 
sources in the same redshift range normalized to have the same
number of sources (blue).
The X-ray sources are not consistent with being drawn from
the same population as the non-X-ray sources. A Mann-Whitney
test gives only a one-sided probability of $4\times10^{-7}$ that
they are the same in the redshift range $z=4.5$--5.5. 
The X-ray sources are, on average, 0.5~mag
redder than the non-X-ray sources in the 1.6~$\mu$m to 1.25~$\mu$m 
flux ratio. They are also more luminous than the non-X-ray
sources, again by about an average 0.5~mag in the 1.6~$\mu$m band.
This would imply that the CANDELS hosts containing high-redshift AGNs 
are more luminous and much dustier than the average galaxy at $z=5$.

However, all of these results are sensitive to the choice of photz 
catalog, and, in particular, to the photometric templates used. 
Indeed, a major concern about the small sample of X-ray detected
$z_{\rm phot}>4.5$ candidates is that the photzs may be
overestimated, in which case the sources are really at lower redshifts 
with lower X-ray luminosities and colors consistent
with the bulk of the galaxies (see Figure~\ref{sant_fs}).

For example, if we instead use the H14 photz
catalog with its mixed galaxy/AGN templates, then where there 
is a H14 photz, most of the X-ray detected CANDELS
$z_{\rm phot}>4.5$ candidates lie at lower redshifts.
ZFOURGE gives photzs that are much closer to the CANDELS 
photzs, but slightly lower, on average. 

In Table~\ref{tab1}, we list all 14 sources 
for which at least one photz catalog gives $z_{\rm phot}>4.5$.
These 14 come either from direct X-ray detections 
(Section~\ref{secdir}) or from NIR/MIR pre-selected X-ray detections.
The NIR/MIR pre-selection re-identifies seven of the eight 
direct X-ray detections. (The exception is L17~\#662,
which lies below the S/N cut of our NIR sample at both 
1.6 and 4.5~$\mu$m).
It also adds six pre-selected X-ray detections, none of which
are in the L17 supplementary catalog. Note that
one of the direct X-ray detections only has a detection 
in the 2--7~keV band (source~14 in Table~\ref{tab1}, or L17~\#802).

At $z>4.5$, we expect that the Ly$\alpha$ forest of the intergalactic
medium should be thick (e.g., Songaila \& Cowie 2010), and we should 
not see flux below the redshifted Ly$\alpha$ wavelength ($<6688$~\AA\
at $z<4.5$).
Vito et al.\ (2018) considered source~6 in Table~1 (L17~\#341) to be clearly
seen at F435W and F606W and hence not at the
$z_{\rm phot}=5.05$ suggested by ZFOURGE. (The measured
F606W flux in the CANDELS catalog is $0.055\pm0.012$ $\mu$Jy). We mark this redshift with 
an `X' in Table~\ref{tab1} and eliminate it from consideration as a 
high-redshift source. We postpone consideration of the optical
properties of the remaining sources to Section~\ref{highz},
where we consider the full SEDs of the candidate high-redshift sources.

\begin{deluxetable*}{ccccccccllcc}
\renewcommand\baselinestretch{1.0}
\tablewidth{0pt}
\tablecaption{X-ray Detected $z_{\rm phot} >4.5$ Candidates\label{tab1}}
\scriptsize
\tablehead{No. & L17 & C18 & R.A. & Decl. & $\log f_{0.5-2}$ & $\log f_{\rm 2-7}$ & G15 $\log f_{0.5-2}$ & \multicolumn{3}{c}{photz} & 850~$\mu$m \\ 
& No. & No. & \multicolumn{2}{c}{(J2000)} & \multicolumn{3}{c}{(erg~cm$^{-2}$~s$^{-1}$)} & C & ZF  & H14 & (mJy)  \\ 
(1) & (2) & (3) & (4) & (5) & (6) & (7) & (8) & (9) & (10) & (11) & (12)}
\startdata
1 & 490 & \nodata & 53.111564 & -27.767771 & -16.01(-17.49) &  -15.26(-16.76) &   -15.91  &  4.52 &      4.73 &  3.24 & 0.00$\pm$0.35\cr
2 & 195 & 45 & 53.040976 & -27.837662 & -16.06(-17.38)  &  -15.45(-16.59) &   -15.96 &     9.73 & \nodata &   4.65 & 2.43$\pm$0.21 \cr
3 & 714 & 7 & 53.158345 & -27.733485 & -16.15(-17.33) &  \nodata (-16.53) &   -16.29 &     5.22 &      3.48 &   2.58 & 5.60$\pm$0.14 \cr
4 & 527 & 72 & 53.119890 & -27.743035 & -16.46(-17.41)&  \nodata  (-16.67)&   -16.48 &     4.84 & \nodata &   2.40 & 1.11$\pm$0.29 \cr
5 & 657 & 17 & 53.146597 & -27.870987 & -16.54(-17.42)  &  -15.40(-16.65) &   -16.38 &     4.70 &      3.57 &   2.47 & 3.80$\pm$0.18 \cr
6 & 341 & \nodata & 53.079375 & -27.741624 & -16.71(-17.36) &  -15.96(-16.59) &    \nodata &   2.09 &      5.05 X  & \nodata & 0.88$\pm$0.37\cr
7 & \nodata & \nodata & 53.087644 & -27.720989 & -16.76(-17.27) &  -16.30(-16.46) &   \nodata &     4.93 & \nodata & \nodata & 1.16$\pm$0.42\cr
8 & 662 & \nodata & 53.147915 & -27.861805  & -16.78(-17.47) &  -16.55(-16.70)  &   \nodata &     4.63 &      4.85  & \nodata & -0.65$\pm$0.34\cr
9 & \nodata & 52 & 53.064687 & -27.862554 & -16.82(-17.37) &  -17.31(-16.61) &   -16.60 &     6.26 & \nodata & \nodata & 1.88$\pm$0.24 \cr
10 & \nodata & \nodata & 53.199966 & -27.774057 & -16.83(-17.32) &  \nodata (-16.54) &  \nodata &     4.95 &      4.42 & \nodata & 0.80$\pm$0.45\cr
11 & \nodata & \nodata & 53.108177 & -27.825122 & -16.83(-17.46) &  \nodata (-16.74) &   \nodata &    4.54 &      4.67 & \nodata & 0.15$\pm$0.26\cr
12 & \nodata & \nodata & 53.197070 & -27.827857 & -16.86(-17.39)  &  -16.68(-16.61)  &    -16.77 &   4.52 &      4.53 & \nodata & 0.86$\pm$0.42\cr
13 & \nodata & \nodata & 53.141126 & -27.764356 & -16.86(-17.46) &  -16.79(-16.75)  &   \nodata &   4.61 &      4.64 & \nodata & 0.97$\pm$0.35\cr
14 & 802 & 54 & 53.181989 & -27.814120 & \nodata (-17.45) &  -16.09(-16.72) &  \nodata &     2.95 &      9.42 & \nodata &  1.82$\pm$0.30 \cr
\enddata
\tablecomments{
Central region direct X-ray detections from L17 and our NIR/MIR pre-selected X-ray detections
(i.e., $>3\sigma$ in at least one of the 0.5--2 or 2--7~keV bands),
for which at least one photz catalog gives
$z_{\rm phot}>4.5$. Table is ordered by decreasing 0.5--2~keV flux. 
Columns: (1) $z>4.5$ candidate source number, 
(2) L17 X-ray catalog number for direct X-ray detections,
(3) C18 ALMA catalog number, when available, 
(4) and (5) NIR R.A. and decl., 
(6) and (7) logarithms of the 0.5--2~keV and 2--7~keV fluxes,
if the source has a positive flux, and logarithms of the rms noise in parentheses,
(8) logarithm of the 0.5--2~keV flux from G15, when available,
(9)--(11) photzs from CANDELS (Santini et al.\ 2015), ZFOURGE (Straatman et al.\ 2016),
and H14, when available (the ZFOURGE photz for source~6 is rejected
and marked with an X in the Table, see text),
and (12) best ALMA 850~$\mu$m flux and rms noise from C18 
(Columns~8 and 9 of their Table~4, when available), 
or SCUBA-2 850~$\mu$m flux and rms noise otherwise.
}
\end{deluxetable*}
\twocolumngrid 

We compare with previous work by G15, who identified 19 NIR pre-selected
X-ray detected CANDELS $z_{\rm phot}>4$ candidates in the central region 
(and a further three outside it) based on the 4~Ms {\em Chandra\/} data set. 
We fail to confirm a number of these---in some cases at a very high significance
level---as did Cappelluti et al.\ (2016), Weigel et al.\ (2015), and Parsa et al.\ (2018). 
Weigel et al.\ (2015) concluded that there were no convincing $z>5$ sources
in the 4~Ms CDF-S sample, while Cappelluti et al.\ (2016) found
14 $z>4$ sources, with possibly 3 at $z>5$. 
Parsa et al.\ (2018) reanalyzed the G15 sample based
on the 4~Ms {\em Chandra\/ image and concluded that there were only 
seven plausible $z>4$ sources, including one at $z>5$.}
Seven Table~\ref{tab1} sources overlap with G15, while another seven
do not. In Column~8 of Table~\ref{tab1},
we give the 0.5--2~keV fluxes from G15
measured using the method described in Fiore et al.\ (2012).
Despite the different methodologies used in the calculations, 
the present values agree well with those  
of G15 for the overlapping sample, though the G15 
values are slightly brighter on average (0.08~dex).
The five sources above $2\times10^{-17}$~erg~cm$^{-2}$~s$^{-1}$
are all common to the two samples and are also in the L17 catalog,
as would be expected given the L17 detection threshold in the
central region. 

We note that all of this previous work was based on the 4~Ms
image, but Giallongo et al.\ (2019) provided a revised list that removed
7 sources from G15 and added 4 based on the deeper 7~Ms
image. Eleven of the sources on this list have $z>4.5$ and lie in the
central region. All are placed at high redshift by one
or other of the photz catalogs, but we only detect
eight above the $3\sigma$ level in the 0.5--2~keV band.
These are the seven overlapping sources in Table~1, plus
source~8 (Luo~\#662). The revised fluxes in Giallongo
et al.\ (2019) are in slightly better agreement with the
present measurements, with an average offset of 0.04~dex.

We note that the remaining three Giallongo et al.\ (2019)
sources have positive measured fluxes in the 
0.5--2~keV band (though not in the 2--7~keV band) with S/N of 1.4, 
1.5, and 2.6.

\begin{figure}[ht]
\centerline{\includegraphics[width=9cm,angle=0]{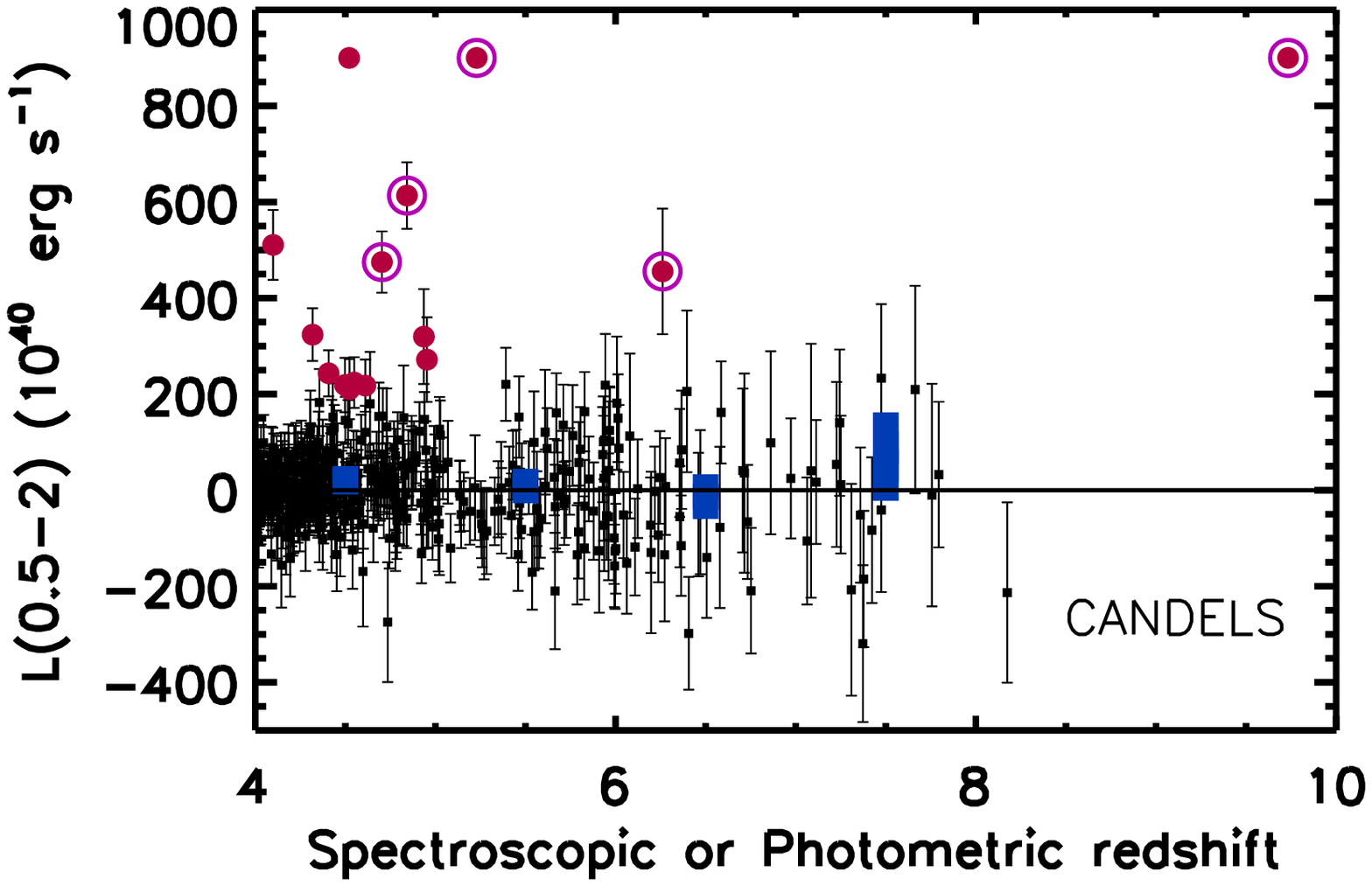}}
\centerline{\includegraphics[width=9cm,angle=0]{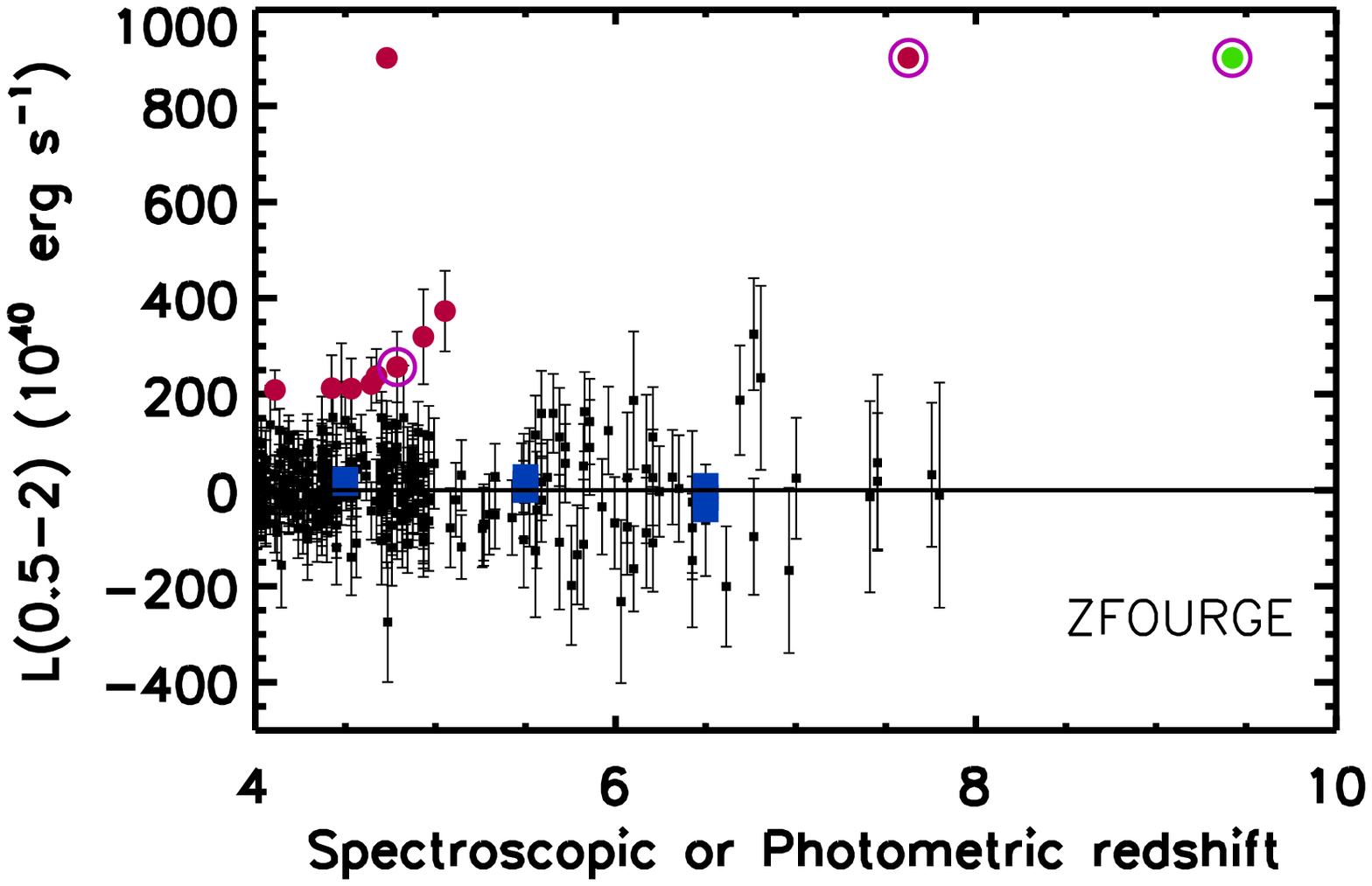}}
\caption{$L_{\rm 0.5--2~keV}$ vs. specz or photz
(CANDELS photzs from Santini et al.\ 2015 in (a) and
ZFOURGE photzs from Straatman et al.\ 2016 in (b))
for the $z>4$ sources (black squares)
in the Guo et al.\ (2013) 1.6~$\mu$m CANDELS catalog that 
lie in the central region and are detected above $10\sigma$ in either
F160W (1.6~$\mu$m) or {\em Spitzer\/} channel~2 ($4.5~\mu$m).
The X-ray error bars are $\pm1\sigma$. 
The red large circles show those sources with $>3\sigma$ detections 
at 0.5--2~keV. Sources with $L_{\rm 0.5-2~keV}>10^{43}$~erg~s$^{-1}$
are shown at $9\times 10^{42}$~erg~s$^{-1}$.
For the one source (L17~\#802) detected in the 2--7~keV band
but not in the 0.5--2~keV band, we use $L_{\rm 2--7~keV}$ (green circle).
(Note that this source only appears in (b), because its CANDELS
photz is only $z_{\rm phot}=2.95$.)
The blue thick lines show the error-weighted mean X-ray luminosities
in the redshift intervals $z=4$--5, 5--6, 6--7, and 7--8,
where the lengths of the lines correspond to $\pm1\sigma$.
Only redshift intervals with more than 10 sources are shown.
The purple enclosing circles mark the sources that are also 
ALMA sources from C18.
\label{sant_ls}
}
\end{figure}

For the present purpose, the key point is that the NIR/MIR pre-selection
finds very few X-ray sources above $z=4.5$, even with the
most optimistic photz estimates. We demonstrate this
in more detail in Figure~\ref{sant_ls}, where we plot $L_{\rm 0.5-2~keV}$
versus redshift using (a) the CANDELS photzs,
which give the largest number of X-ray detected $z_{\rm photz}>4$ candidates
(red circles), and (b) the ZFOURGE photzs.

Many of the X-ray detected sources are low enough in 
luminosity that they could contain substantial star formation 
contributions to the X-ray luminosities (see Barger et al.\ 2019 for
a recent discussion). These contributions may become significant at or below 
$\sim 10^{42}~$erg~s$^{-1}$, though for extreme star formers, the
values produced by the hard X-ray binaries 
can be somewhat higher, with star formation producing as much as
$10^{42.5}~$erg~s$^{-1}$ for strong submillimeter sources.

Only one of the high-redshift candidates (source~2 in Table~1,
C18~\#45 or L17~\#195) is detected in the 1.4~GHz sample of Miller
et al.\ (2013), with a flux of 89.2~$\mu$Jy. This is only a moderately bright
submillimeter source with an 850~$\mu$m flux of $2.4\pm0.2$~mJy.
Regardless of the chosen photz, the low ratio of the submillimeter
flux to the radio power would argue that the radio emission
is AGN dominated (Barger et al.\ 2017).

We also show the error-weighted 
mean X-ray luminosities on Figure~\ref{sant_ls}, where the lengths of the 
lines correspond to $\pm1\sigma$ (blue thick lines). We summarize these 
in Table~\ref{tab2} for the CANDELS photzs. 
Above $z=5$, there is no significant signal at 
a very low level, consistent with previous results
(Willott 2011; Cowie et al.\ 2012; Vito et al.\ 2016).
Thus, there is no evidence of any AGN activity in the population
as a whole. Similar results are found for the other photz estimates.

We also mark the sources that are also ALMA sources from 
Cowie et al.\ (2018; hereafter, C18; their Table~4) with purple enclosing circles.
These six sources are identifiable in Table~\ref{tab1} by their C18 numbers
in Column~2. In both panels, all of the luminous X-ray sources above $z=5$ are 
ALMA sources. This allows us to make an alternative estimate of the
photzs using the FIR SEDs. Some of these
sources could be luminous dusty star formers, but, as we discuss
in Section~\ref{alma_select}, these alternative FIR SED redshift estimates 
(hereafter, FIRzs) suggest that at least some of the sources that the optical/NIR 
photzs place at very high redshifts are, in fact, at lower redshifts.

\begin{figure}[ht]
\centerline{\includegraphics[width=9cm,angle=0]{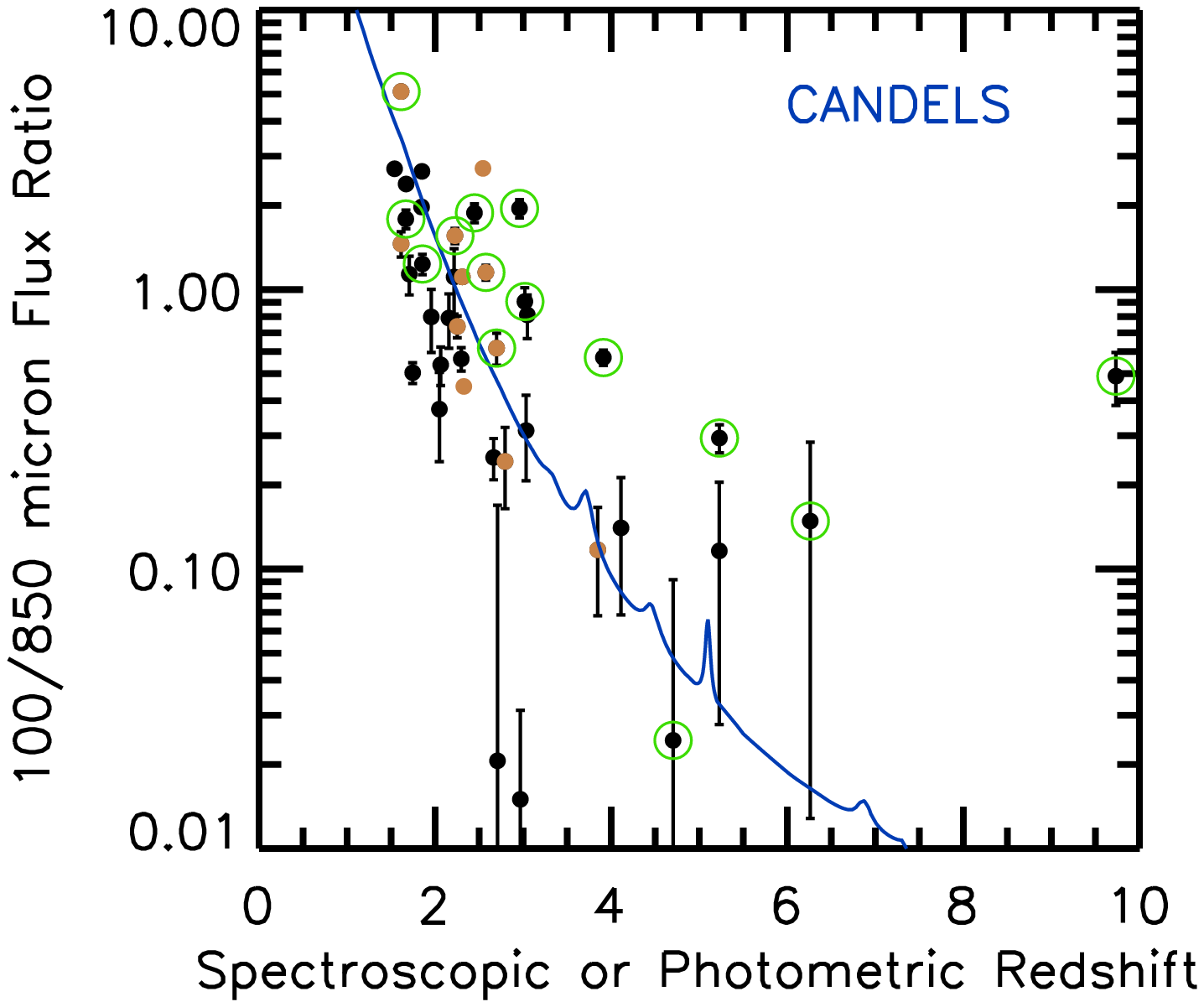}}
\centerline{\includegraphics[width=9cm,angle=0]{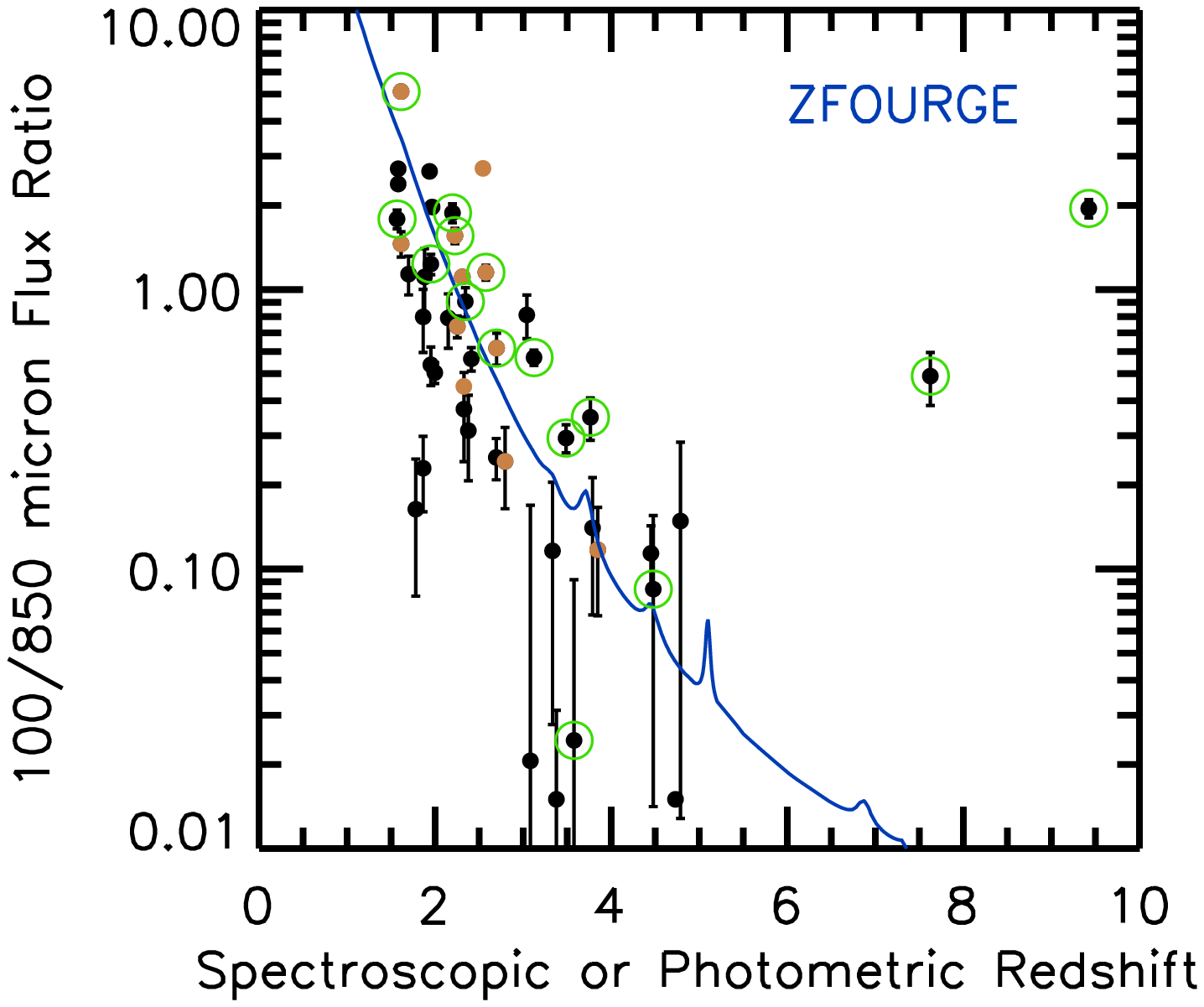}}
\caption{100 to 850~$\mu$m flux ratio vs. redshift 
for the 54 ALMA sources in the central region
with 850~$\mu$m fluxes above 1.65~mJy and NIR/MIR counterparts
that are included in our NIR sample.
The gold circles show sources with speczs and the black
circles those with photzs. The photzs used in (a) are from
CANDELS (Santini et al.\ 2015), while those used in (b)
are from ZFOURGE (Straatman et al.\ 2016). 
The error bars are $\pm1\sigma$.  The blue
curve shows the flux ratio for a redshifted Arp~220 SED template. 
The green enclosing circles show sources where either $L_{\rm 0.5-2~keV}$ 
or $L_{\rm 2-7~keV}$ are $>10^{42.5}$~erg~s$^{-1}$.
\label{fir_color}
}
\end{figure}

\subsection{ALMA Pre-Selection}
\label{alma_select}
In C18, we presented our SCUBA-2 and ALMA observations of the CDF-S. 
The SCUBA-2 data are very deep in the inner 100~arcmin$^2$,
typically with an 850~$\mu$m flux error less than 0.4~mJy,
but degrade with off-axis angle, reaching a value of 1.3~mJy at a $10'$ outer
radius. The maximum noise within the X-ray central region
(i.e., within an off-axis angle of $5\farcm7$; see Section~\ref{secdir})
is 0.51~mJy, and here all of the $>2.25$~mJy
850~$\mu$m SCUBA-2 sources have been observed with ALMA in band 7.
(We refer to all the submillimeter fluxes as 850~$\mu$m,
ignoring the small differences in the ALMA band 7 wavelength centers.)
A number of fainter SCUBA-2 sources with 850~$\mu$m fluxes
below 2.25~mJy have also been observed, along with a small number of 
serendipitous ALMA sources extending to fluxes less than 1~mJy.

After restricting the area of the individual ALMA images to the 
half-power radius of $8\farcs75$ where detections are robust, the combined
ALMA imaging covers a total area of 7.2~arcmin$^2$, with most
of that area (i.e., 5.9~arcmin$^2$) concentrated in the central region.
In total, C18 found 75 $>4.5\sigma$ ALMA sources.
Here we define our ALMA sample as the 58 ALMA sources in the central
region that lie 
above the SCUBA-2 confusion limit of 1.65~mJy (Cowie et al.\ 2017).
We measured the X-ray fluxes at the positions of this sample as described
in Section~\ref{xray_preselect}.

\begin{deluxetable}{cccc}
\renewcommand\baselinestretch{1.0}
\tablewidth{0pt}
\tablecaption{0.5--2~keV Luminosity\label{tab2}}
\tablehead{$z_{\rm min}$ & $z_{\rm max}$ & Number & Mean Luminosity \\
& & & ($10^{40}$~erg~s$^{-1}$) \\
(1) & (2) & (3) & (4)  }
\startdata
    4.0  &     5.0  &       295  &   20.7$\pm$3.1 \cr 
   5.0  &      6.0  &        75  &   7.7$\pm$9.8  \cr
    6.0  &     7.0  &       32  &   -8.0$\pm$20.3 \cr 
    7.0  &     8.0  &       18  &   70.1$\pm$39.0 \cr 
\enddata
\tablecomments{
Error-weighted mean 0.5--2~keV luminosities for the $z_{\rm phot}>4$
sources using the CANDELS photzs (Santini et al.\ 2015). 
Columns: 
(1) and (2) Minimum and maximum redshift of bin, 
(3) number of sources in bin, 
and (4) error-weighted mean X-ray luminosity of sources
in bin.
}
\end{deluxetable}

\begin{deluxetable*}{cccccccccccc}
\renewcommand\baselinestretch{1.0}
\tablewidth{0pt}
\tablecaption{ALMA Pre-Selected X-ray Detected $z>4.5$ Candidates\label{tab3}}
\scriptsize
\tablehead{Table~1 & C18 & L17 & R.A. & Decl. & $\log f_{\rm 0.5-2}$  & $\log f_{\rm 2-7}$ & \multicolumn{3}{c}{photz} & FIRz  & 850~$\mu$m \\ No. & No. & No. & \multicolumn{2}{c}{(J2000)} & 
\multicolumn{2}{c}{(erg~cm$^{-2}$~s$^{-1}$)} & C & ZF & H14 & C18 & (mJy)  \\ (1) & (2) & (3) & (4) & (5) & (6) & (7) & (8) & (9) & (10) & (11) & (12)}
\startdata
3 & 7 & 714 & 53.1584 & -27.7336 & -16.15(-17.32) &  \nodata(-16.53) &      5.22 &      3.48 &   2.58 &  3.37 &  5.60$\pm$0.14\cr
5 & 17 & 657 & 53.1466 & -27.8710 & -16.54(-17.41) &  -15.40(-16.65) &      4.70 &      3.57 &   2.47 &  7.99 &  3.80$\pm$0.18\cr
\nodata & 19 & 472 & 53.1088 & -27.8690 & \nodata(-17.44) &  -16.03(-16.68) & \nodata &      4.47 & \nodata &  6.69 &  3.62$\pm$0.17\cr
\nodata & 44 & \nodata & 53.0872 & -27.8402 & -16.74(-17.49) &  -17.06(-16.77) & \nodata & \nodata & \nodata &  5.26 &  2.21$\pm$0.12\cr
2 & 45 & 195 & 53.0411 & -27.8377 & -16.06(-17.38) &  -15.45(-16.59) &      9.73 &      7.62 &   4.65 &  3.09 &  2.20$\pm$0.23\cr
9 & 52 & \nodata & 53.0648 & -27.8626 & -16.82(-17.37) &  -17.31(-16.61) &      6.26 &      4.78 & \nodata &  3.63 &  1.88$\pm$0.24\cr
14 & 54 & 802 & 53.1820 & -27.8142 & \nodata(-17.44) &  -16.09(-16.71) &      2.95 &      9.42 & \nodata &  1.85 &  1.82$\pm$0.30\cr
\enddata
\tablecomments{
Central region ALMA pre-selected X-ray detected sources 
($>3\sigma$ in at least one of the 0.5--2 or 2--7~keV bands) for which at least one 
photz catalog gives $z_{\rm phot}>4.5$. 
Table is ordered by decreasing 850~$\mu$m flux.
Columns: (1) Table~\ref{tab1} source number, (2) C18 ALMA catalog number, 
(3) L17 X-ray catalog number, when available,
(4) and (5) ALMA R.A. and decl., 
(6) and (7) logarithms of the 0.5--2~keV and 2--7~keV fluxes, 
if the source has a positive flux, and logarithms of the rms noise in parentheses,
(8)--(10) photzs from CANDELS (Santini et al.\ 2015), ZFOURGE (Straatman et al.\ 2016),
and H14, when available,
(11) FIRz from C18,
and (12) best ALMA 850~$\mu$m flux and rms noise from C18 (Columns~8 and 9 of their Table~4).
}
\end{deluxetable*}
\twocolumngrid


Most of our ALMA sample have NIR/MIR counterparts
(see Table~5  and Figure~10 of C18), and many of these are
strong detections, so they are 
already included in our NIR sample of Section~\ref{nir_select}
(54 of the 58 ALMA sources).
For these sources, the FIR/submillimeter SEDs provide an independent check
on the optical/NIR/MIR photzs. 

While full fits to the FIR/submillimeter SEDs are optimal, and we turn to these 
for the high-redshift candidates below, we can most
easily visualize the constraints provided by FIR/submillimeter
data with a simple color plot versus redshift.
In Figure~\ref{fir_color}, we plot the 100/850~$\mu$m flux ratio 
versus redshift (gold circles for speczs, black circles for photzs)
for the 54 sources. We note that all fluxes are corrected to
total and appear to match well (see C18). The 850~$\mu$m
flux increases relative to the 100~$\mu$m flux with increasing
redshift, resulting in a rapidly dropping ratio. We illustrate this
by redshifting the Arp~220 SED template (blue curve). 
The CANDELS photzs in Figure~\ref{fir_color}(a) have an
apparently high upward scatter relative to the Arp~220 
ratio, while the ZFOURGE photzs in Figure~\ref{fir_color}(b)
lie close to the blue
curve, with the exception of the two very high-redshift candidates.
In both panels, the luminous X-ray sources (green enclosing circles)
tend to lie high, while the small number of high-redshift candidates lie
very high. Figure~\ref{fir_color} suggests that for these sources, we are
either significantly overestimating their photzs, or their
FIR SEDs are highly anomalous.

\begin{figure}[ht]
\centerline{\includegraphics[width=9cm,angle=0]{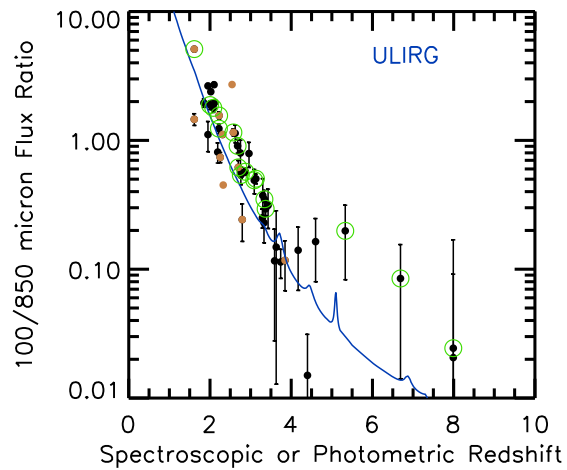}}
\caption{100 to 850~$\mu$m flux ratio vs. redshift for our ALMA 
sample in the central region with 850~$\mu$m fluxes above 1.65~mJy.
The gold circles show sources with speczs and the black
circles those with FIRzs. The FIRzs are from C18.
The error bars are $\pm1\sigma$.  
The blue curve shows the flux ratio for a redshifted Arp~220 SED
template. The green enclosing circles show sources where either 
$L_{\rm 0.5-2~keV}$ or $L_{\rm 2-7~keV}$ are $>10^{42.5}$~erg~s$^{-1}$.
\label{fir_firz}
}
\end{figure}

The four remaining sources in our ALMA sample 
(i.e., those not already in our NIR sample)
either have no NIR/MIR counterpart, are blended with a brighter 
neighbor object in the optical/NIR, or lie outside the
CANDELS field (see Figure~10 of C18).
For these sources, we can only rely on the fit to the FIR SED. 

As described in C18, we fitted the full
ALMA sample using an Arp~220 SED template to obtain FIRz
estimates (Table~5 of C18), since Arp~220 generally provides a 
good approximation to the SEDs of ULIRGS with high speczs. 

In Figure~\ref{fir_firz}, we show the same simple color plot,
this time for the FIRzs from C18. Since the
FIRz fitting depends strongly on the 100/850~$\mu$m flux ratio,
the FIRzs are tightly correlated with this ratio. This only
breaks down at the faintest 100~$\mu$m fluxes, where the 
ratio becomes noisy and the SED fit is more constrained 
by longer wavelength data.

The primary concern is the possibility that this may not be 
representative of sources containing AGNs. There is one known X-ray 
source with $z_{\rm spec}>4$ in the wider CDF-S field. This is the
Compton-thick source found by Gilli et al.\ (2011). 
We note the recent search by Circosta et al.\ (2019) for obscured
AGNs in the CDF-S and their analysis of their SEDs, but only
the Gilli et al.\ source lies at $z>4$.

\begin{figure}[ht]
\centerline{\includegraphics[width=9cm,angle=0]{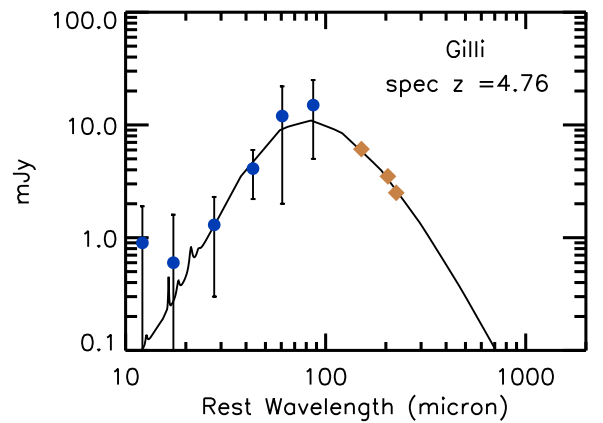}}
\caption{FIR SED for the $z=4.76$ dusty X-ray source
in the outer part of the CDF-S. Data are from Gilli et al.\ (2014):
{\em Herschel\/} data are shown in blue, and ALMA in gold. 
The error bars are $\pm1\sigma$. The black curve shows the Arp~220 
SED template fit with $z_{\rm FIR}=4.77$.
\label{gilli}
}
\end{figure}

While the Gilli et al.\ (2011)
source is outside the central region at an off-axis angle of $8'$,
it has a good specz, it is strongly detected in the submillimeter, and it has
been intensively observed with ALMA. Thus, it appears to be a
near ideal source to compare with the present sample.
It, too, is well approximated by an Arp~220 SED template, as we
illustrate in Figure~\ref{gilli}, and $z_{\rm FIR}=4.77$
is extremely close to $z_{\rm spec}=4.76$,
as are the photz estimates ($z_{\rm phot}=4.48$ from CANDELS,
$z_{\rm phot}=4.84$ from ZFOURGE, and $z_{\rm phot}=4.69$ from 
H14). Thus, for this object, the various fitting procedures work well 
and are consistent. While results on one source are not proof that the 
FIRz estimates are reliable in all cases, we further note that we find little 
difference between the residual fits in AGN and non-AGN ALMA sources, 
implying that there are no apparent systemic differences.

In Table~\ref{tab3}, we summarize all of the ALMA pre-selected
X-ray detected
sources above 1.65~mJy where one or more of the photzs or FIRzs
place the source at $z>4.5$.
(This may be compared with Table~6 in C18, which gives
all of the ALMA sources above 1.65~mJy where 
either the ZFOURGE photz or the FIRz places the source at $z>4$.)
Column~1 lists the number in this paper's Table~\ref{tab1},
when available, and Column~2 the number from C18's Table~4.
The remaining columns give the L17 catalog number,
the right ascensions and declinations, the X-ray fluxes, the various 
redshift estimates, and the ALMA 850~$\mu$m flux. 

The FIRz fits place three of Table~\ref{tab3}'s sources at $z_{\rm FIR}>4.5$.
One of these sources (C18~\#17 or L17~\#657; $z_{\rm FIR}=7.99$)
is placed at $z_{\rm phot}=3.6$ by ZFOURGE and at 
$z_{\rm photz}=4.7$ by CANDELS. Another source
(C18~\#19 or L17~\#472; $z_{\rm FIR}=6.69$) 
is not included in Table~\ref{tab1},
because the one available photz by ZFOURGE ($z_{\rm phot}= 4.47$) 
places it below $z=4.5$. A third source
(C18~\#44; $z_{\rm FIR}=5.26$) is not in either the main or
supplemental L17 catalogs, but we detect it in the
0.5--2~keV band at the $5.6\sigma$ level. It is not significantly
detected in the 2--7~keV band.
However, it is is too faint in the optical/NIR 
to have an optical/NIR photz and hence also does
not appear in Table~\ref{tab1}. 

Combining the samples of Tables~\ref{tab1} 
and \ref{tab3} gives a total of 16 X-ray detected $z>4.5$ candidates,
of which eight are also ALMA 
detected\footnote{Source~4 in Table~\ref{tab1} is the eighth ALMA 
source; it is not in Table~\ref{tab3}, because it is below 1.65~mJy.}.

\section{Individual High-Redshift AGN Candidates}
\label{highz}
As can be seen from Section~\ref{other_select}, there is no consistency
in the selection of X-ray detected high-redshift candidates between the 
various photz and FIRz estimates. In particular, the FIRzs often do not support
the high-redshift candidates identified by the optical/NIR photzs, while
the photzs would place some of the high-redshift 
candidates identified by FIRzs at lower redshifts. 
However, there are substantial uncertainties
in the template fitting, particularly for the FIRz fits, and
we must also worry about AGN contributions to the templates
for both the photz and FIRz fits. In this section, we
try to assess what the most robust constraints are on
the small number of X-ray detected high-redshift 
candidates found by any of the methods.

\begin{figure*}
\includegraphics[width=2.8in,angle=0]{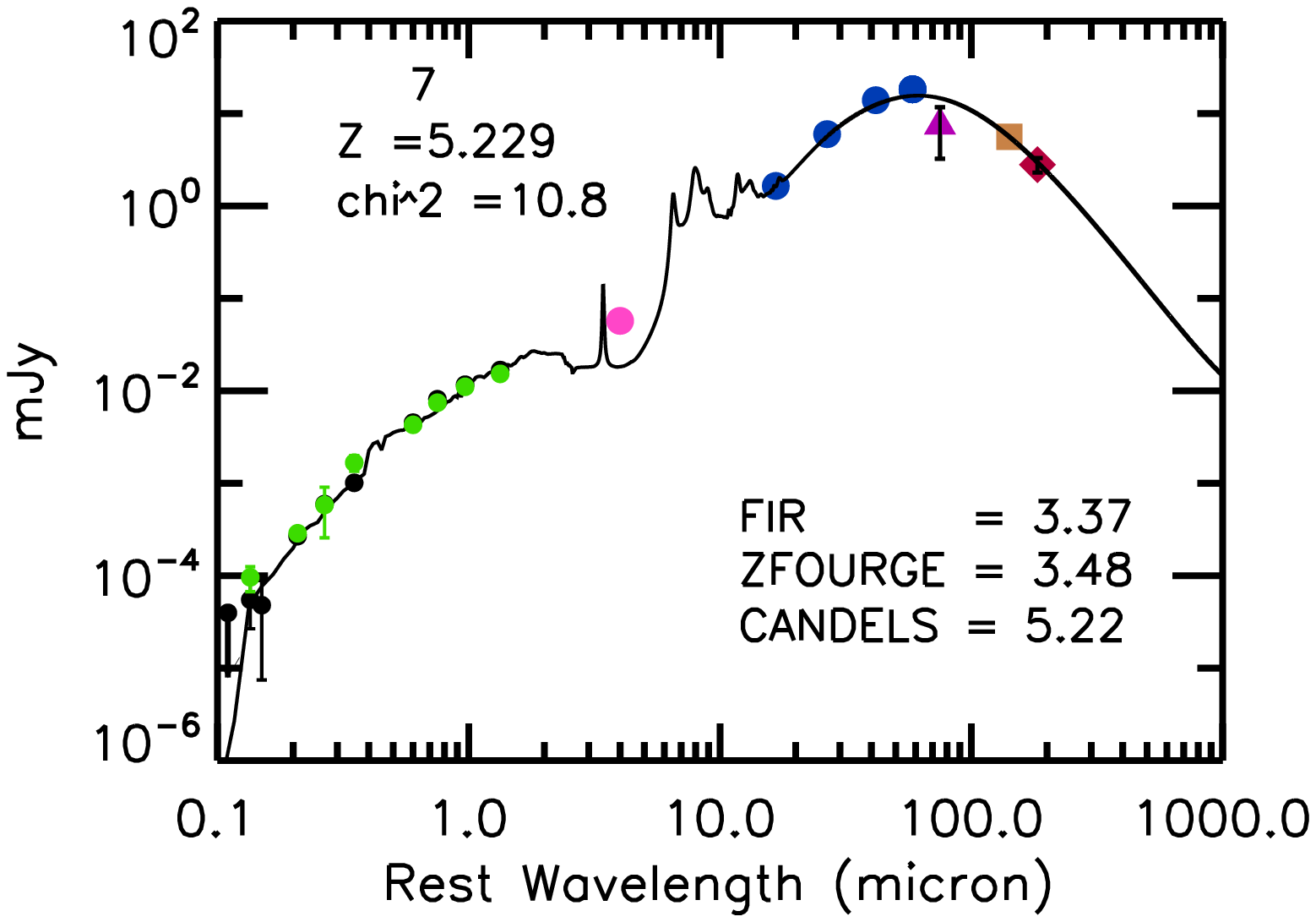} 
\includegraphics[width=2.8in,angle=0]{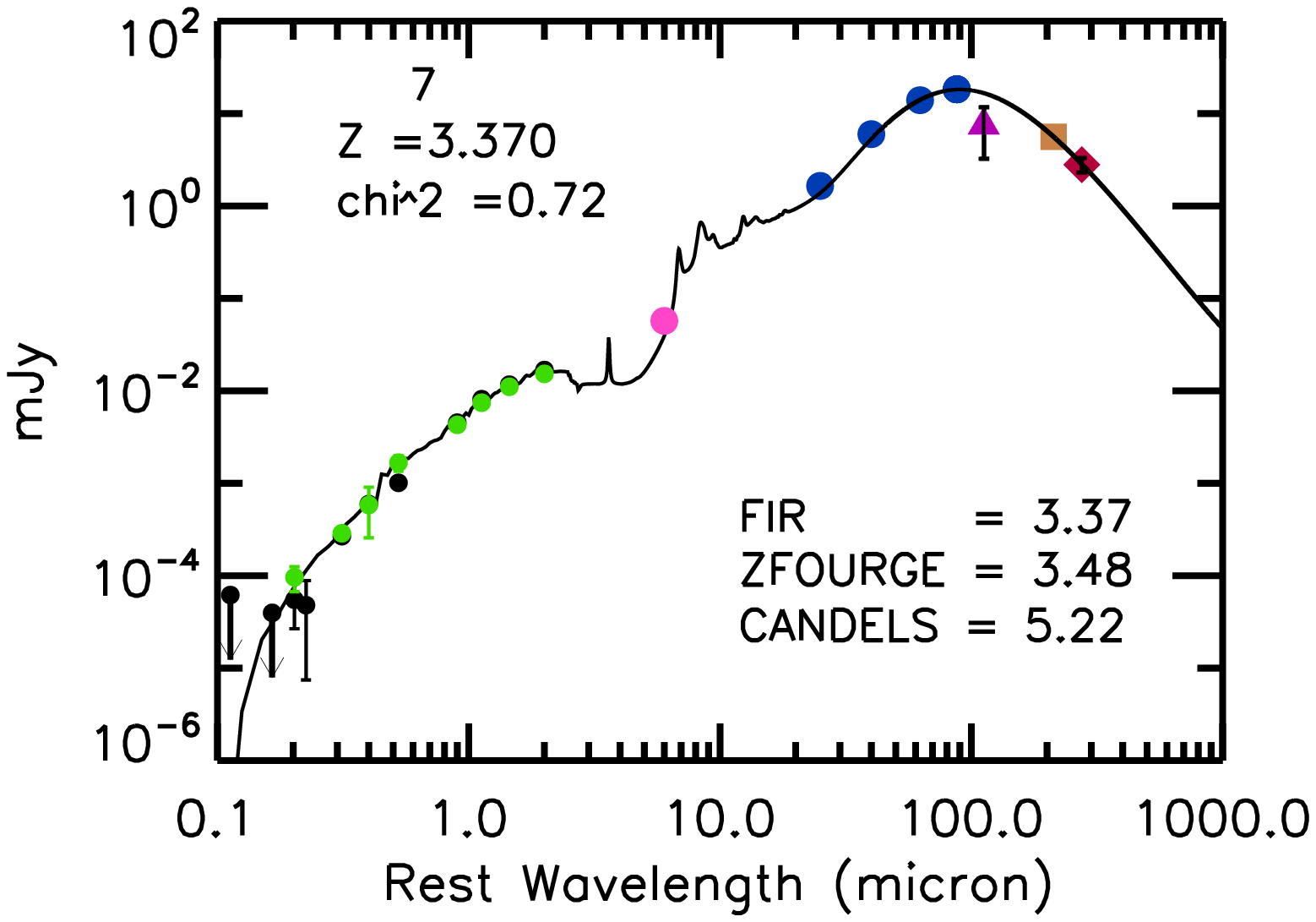} 
\includegraphics[width=2.8in,angle=0]{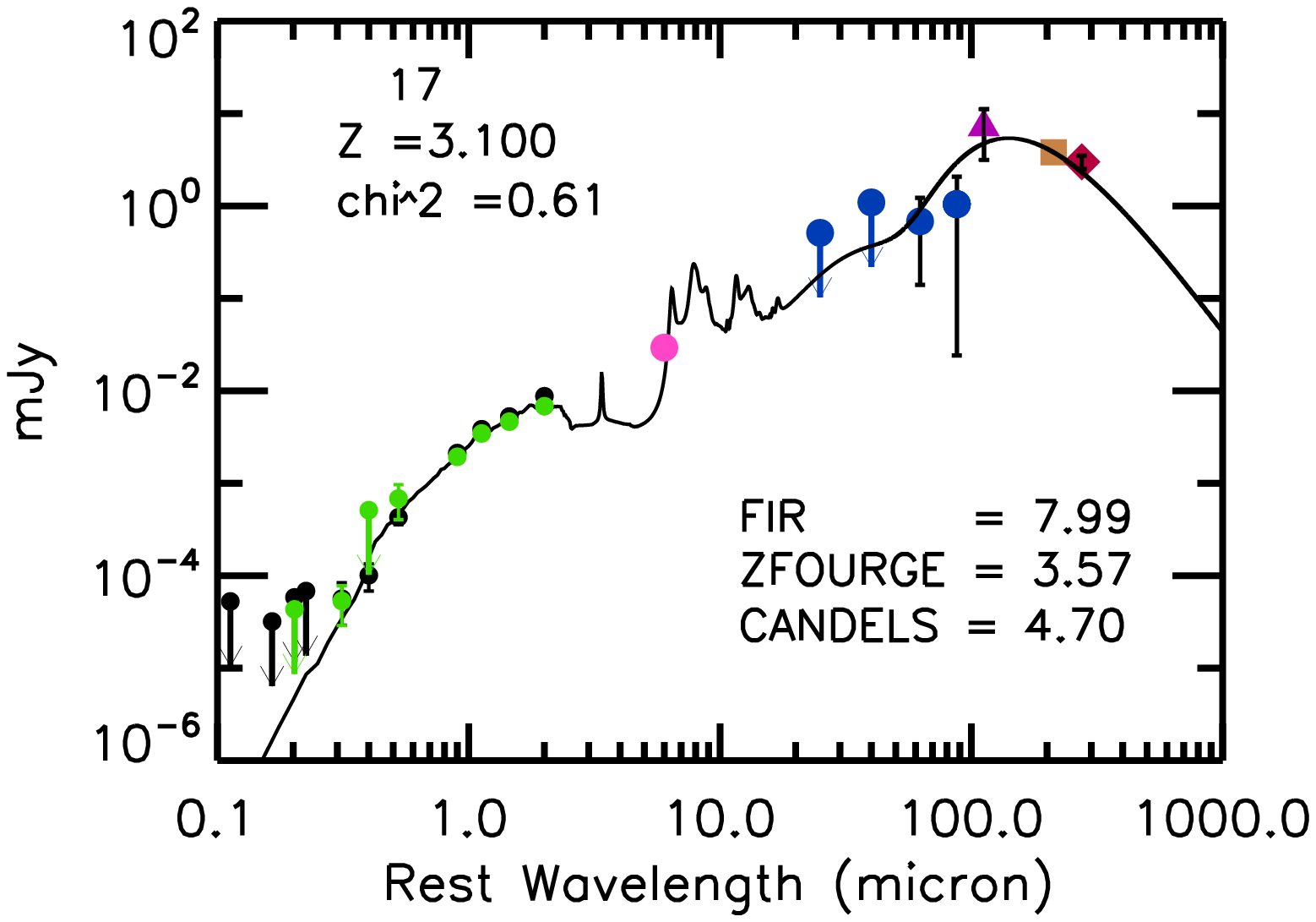} 
\includegraphics[width=2.8in,angle=0]{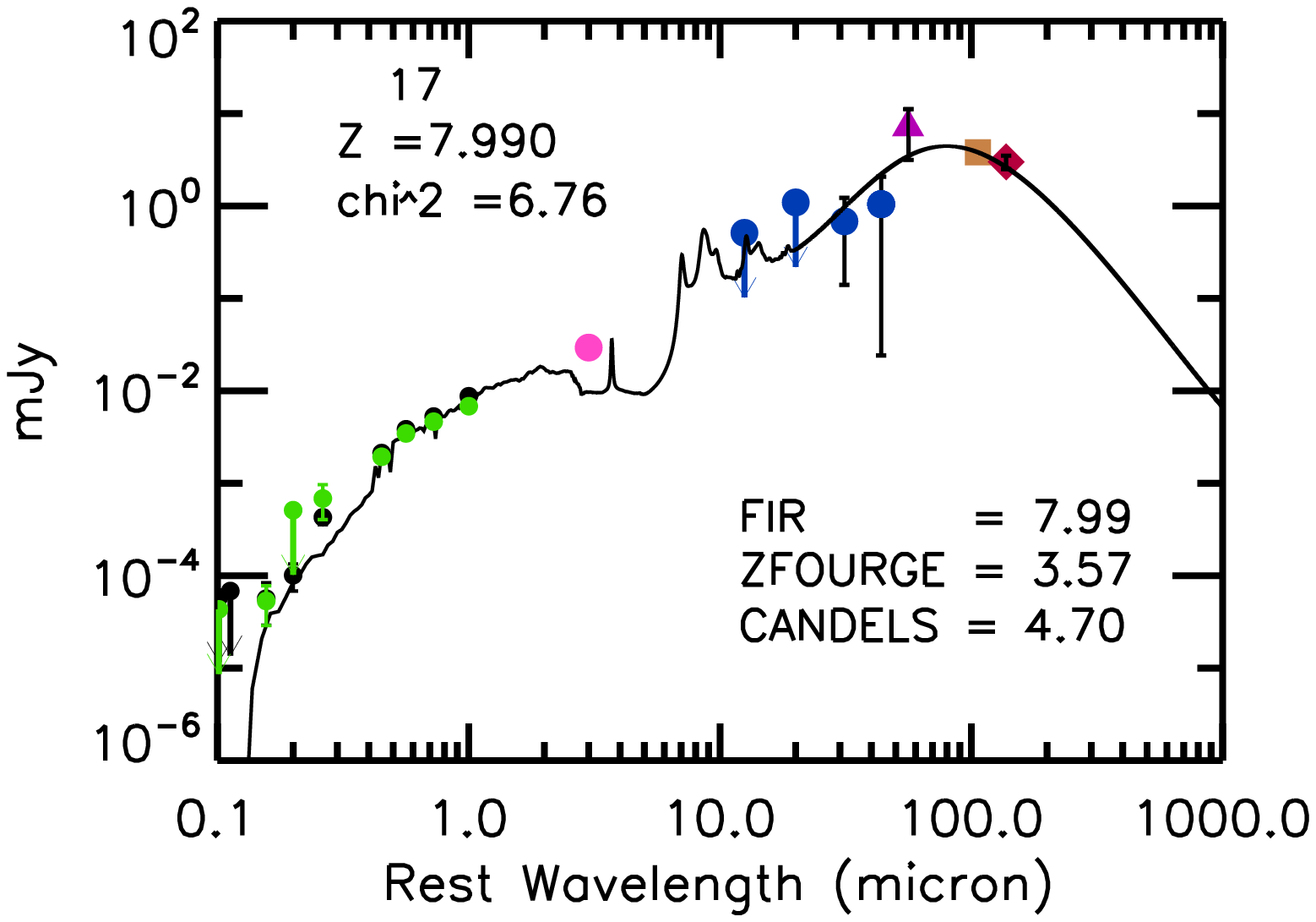} 
\includegraphics[width=2.8in,angle=0]{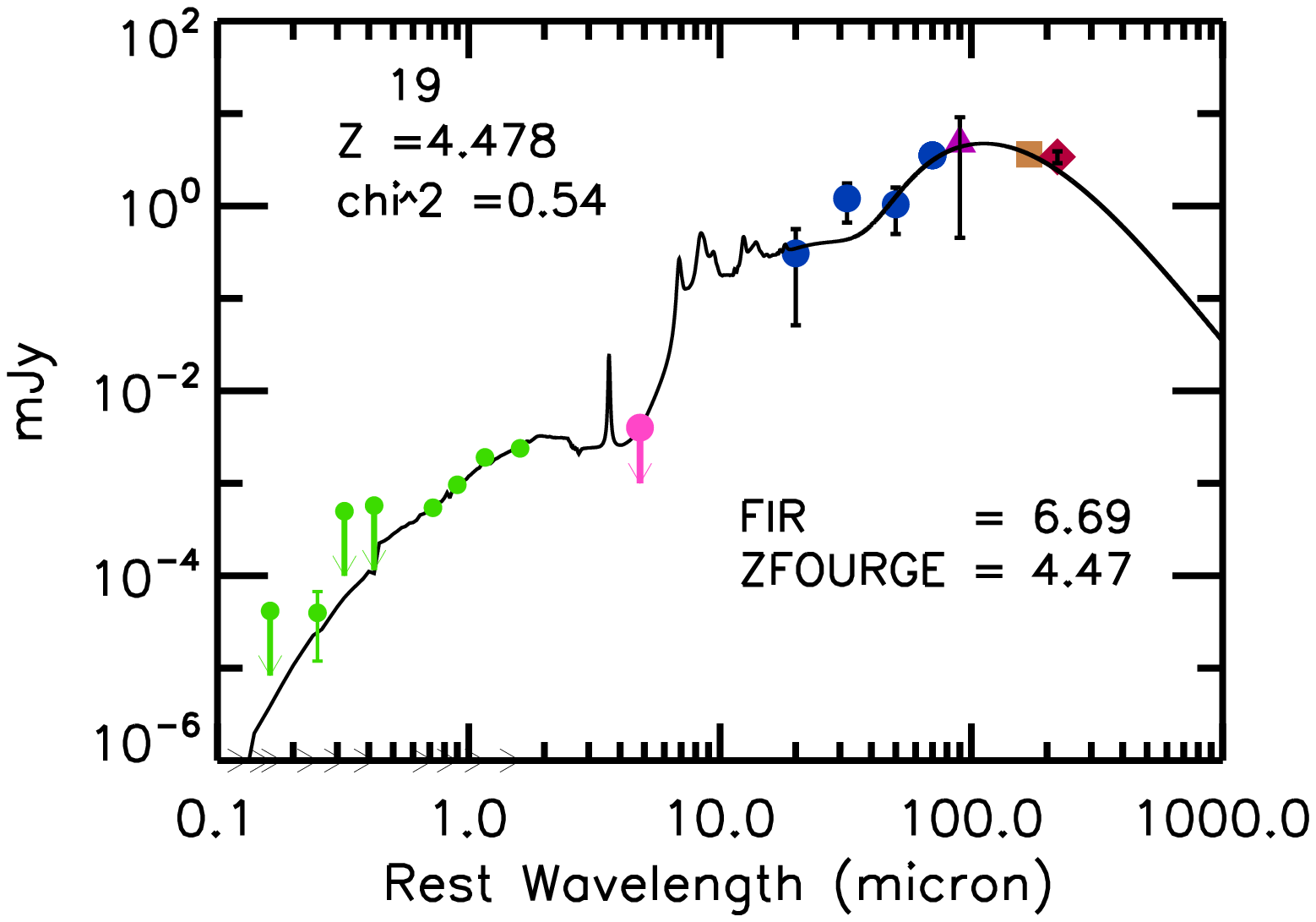} 
\includegraphics[width=2.8in,angle=0]{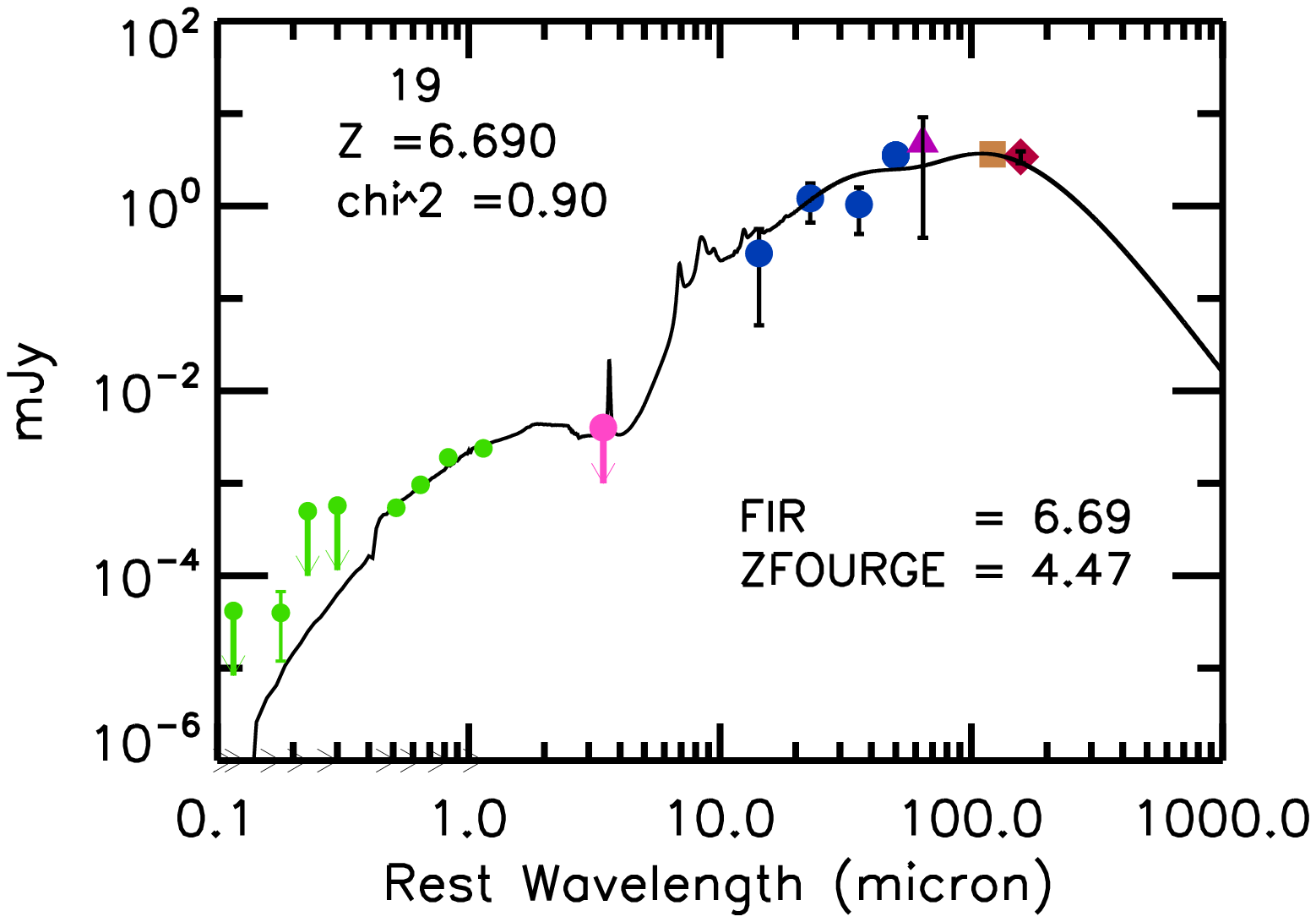} 
\includegraphics[width=2.8in,angle=0]{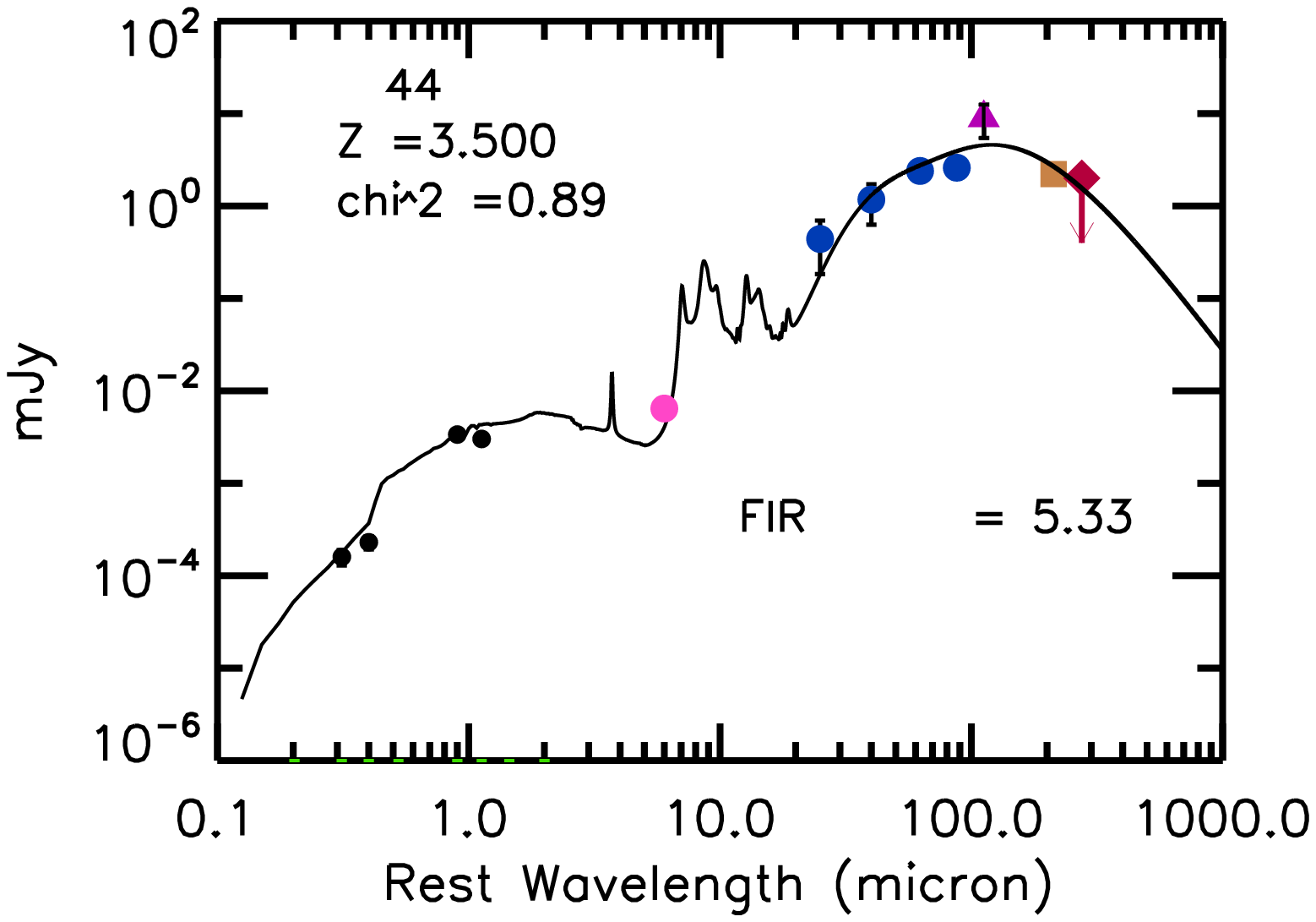} 
\hspace{3.8cm}\includegraphics[width=2.8in,angle=0]{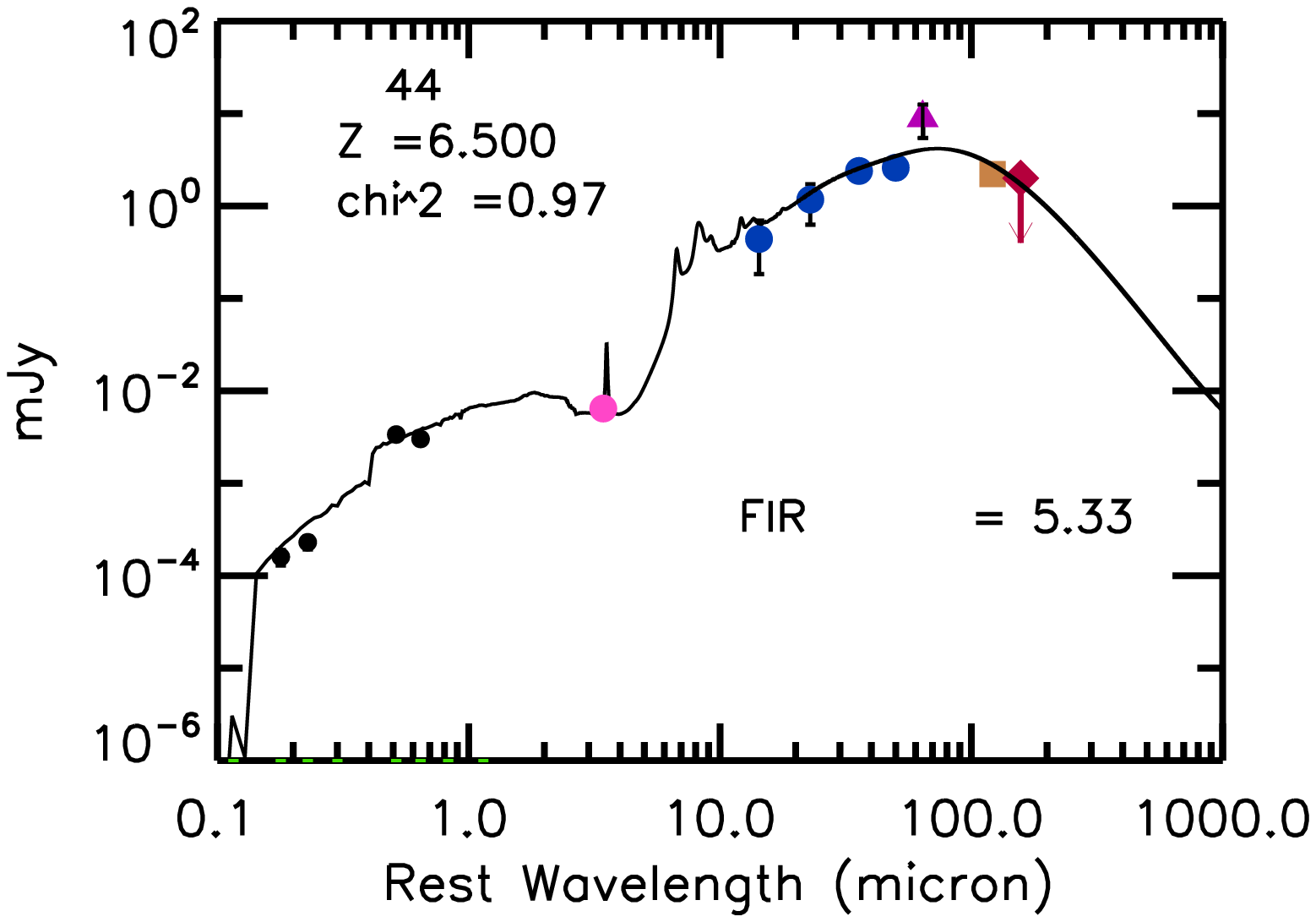} 
\caption{The best-fit MAGPHYS model SED fit (black curve) overlaid
on the data points for the ALMA pre-selected X-ray detected 
high-redshift candidates (green = ZFOURGE, black = CANDELS,
pink = {\em Spitzer\/} 24~$\mu$m, blue = {\em Herschel\/}, purple = SCUBA-2 450~$\mu$m,
gold = ALMA 850~$\mu$m, red = ALMA or AzTEC 1.1~mm; error bars are
$1\sigma$, and sources with $<1\sigma$ detections  are shown
as $2\sigma$ upper limits with downward pointing arrows.
In each case, the left panel shows the model SED fit using either the 
CANDELS or ZFOURGE photz, and the right panel that using the FIRz.
The only exceptions are source~17, where we use $z=3.1$ in the left 
panel (see text for details), 
and source~44, where there is no optical/NIR photz, 
and we use $z=3.5$ in the left panel and $z=6.5$ in the right panel
(see text for details).
The numbers in the top-left corner of each panel are the 
source numbers from C18, the chosen redshift for the 
SED fit, and the $\chi^2$ from the MAGPHYS code.
\label{wide_sedz}
}
\end{figure*}

\begin{figure*}
\figurenum{8}
\includegraphics[width=2.8in,angle=0]{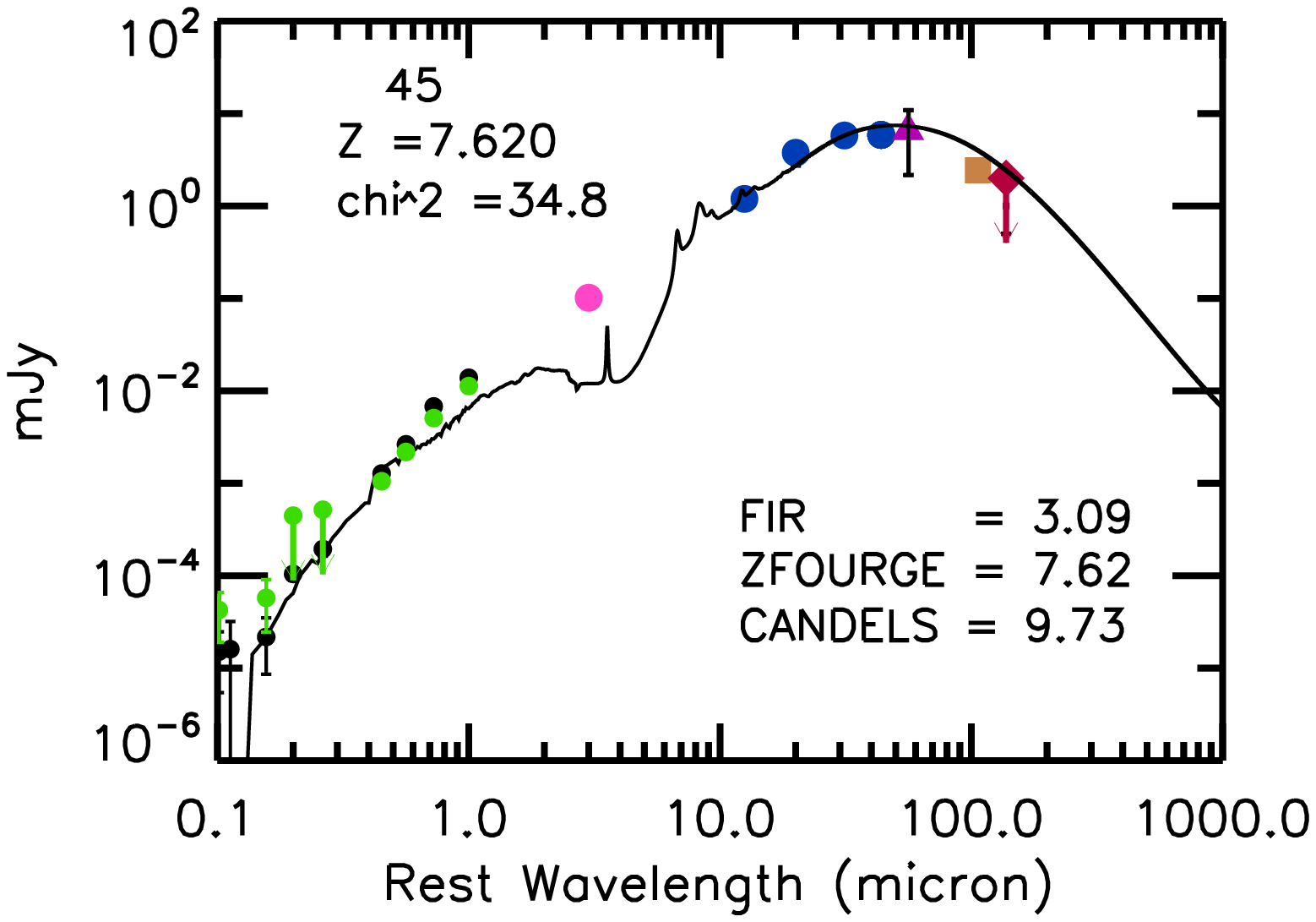} 
\includegraphics[width=2.8in,angle=0]{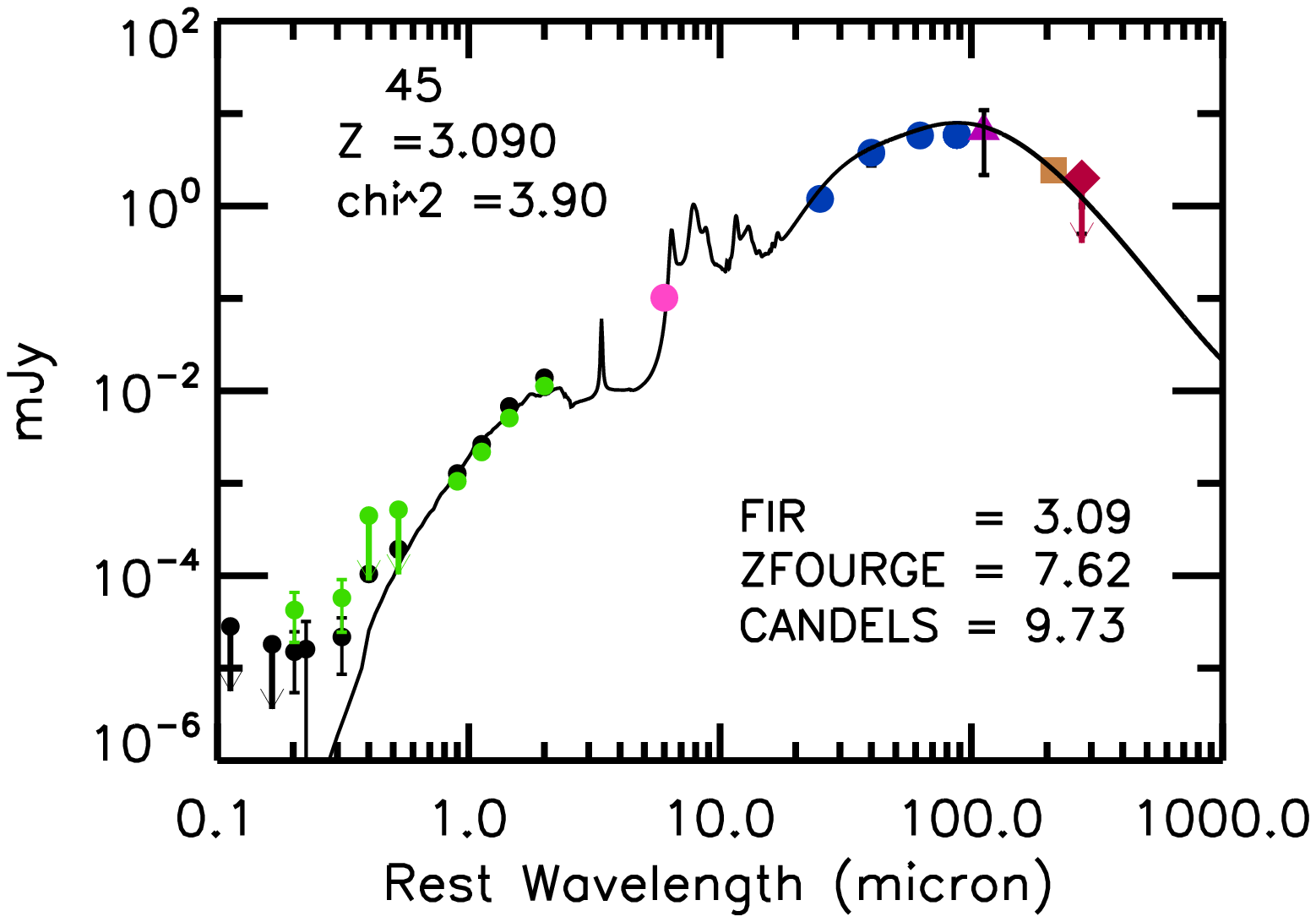} 
\includegraphics[width=2.8in,angle=0]{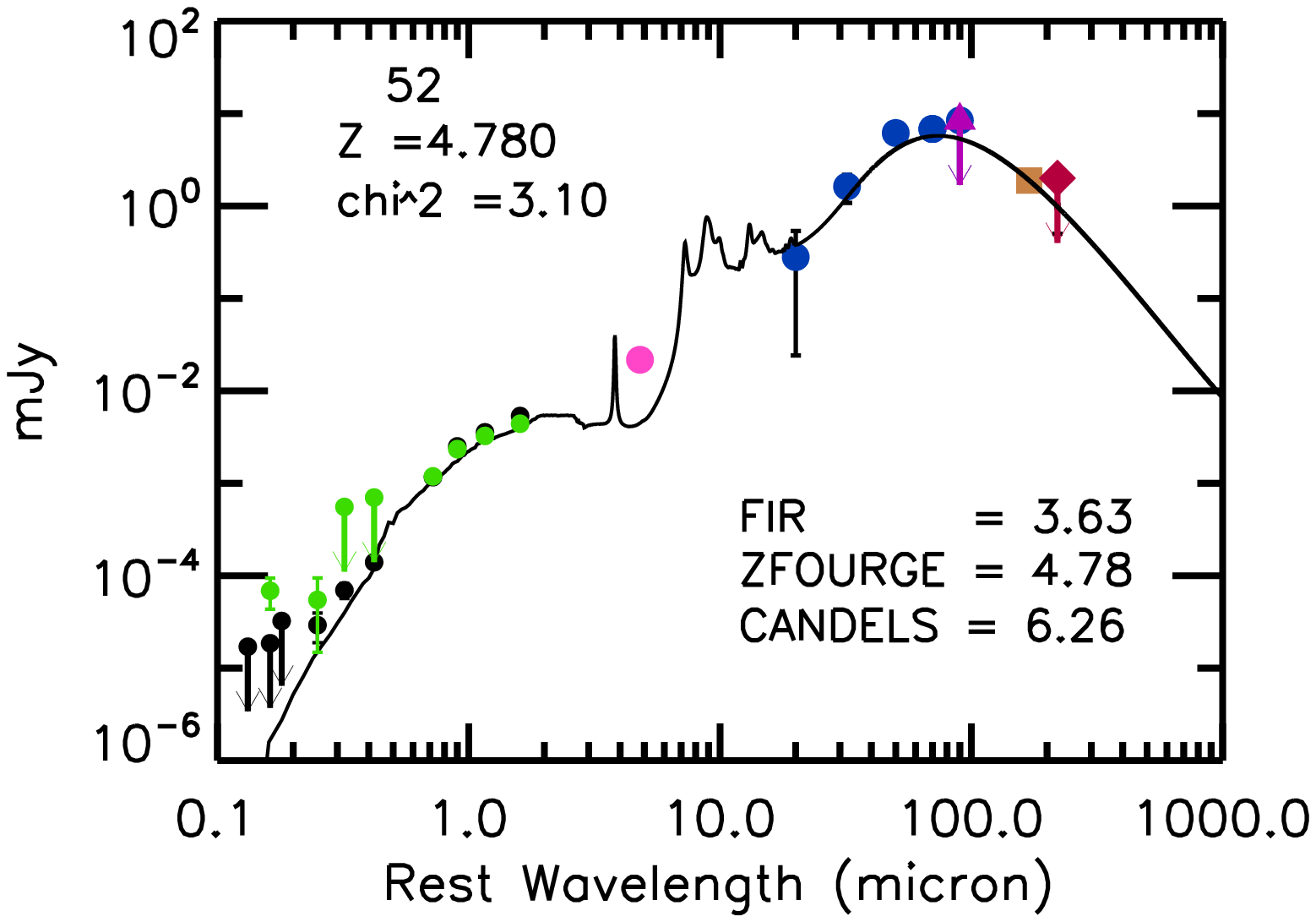} 
\includegraphics[width=2.8in,angle=0]{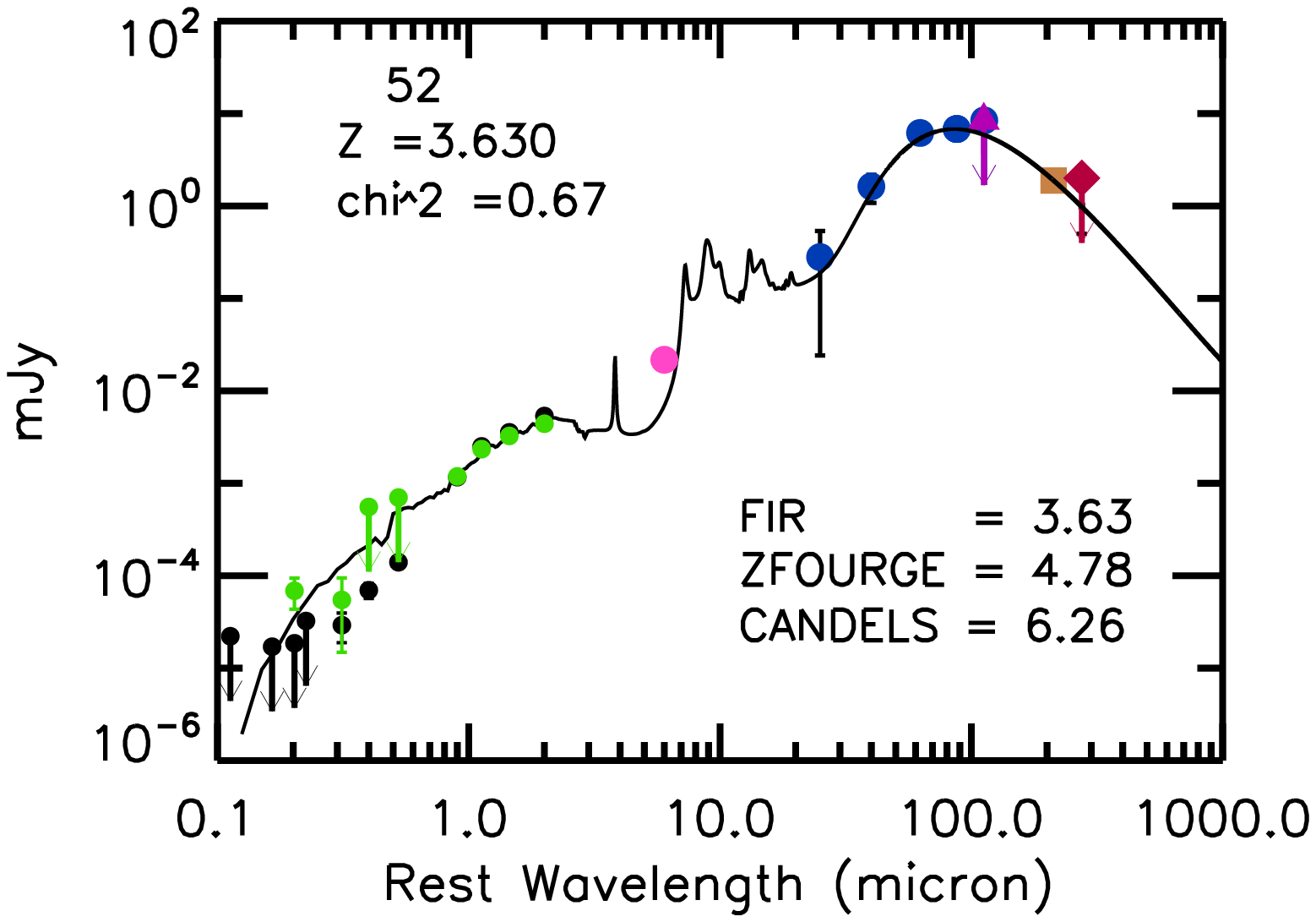} 
\includegraphics[width=2.8in,angle=0]{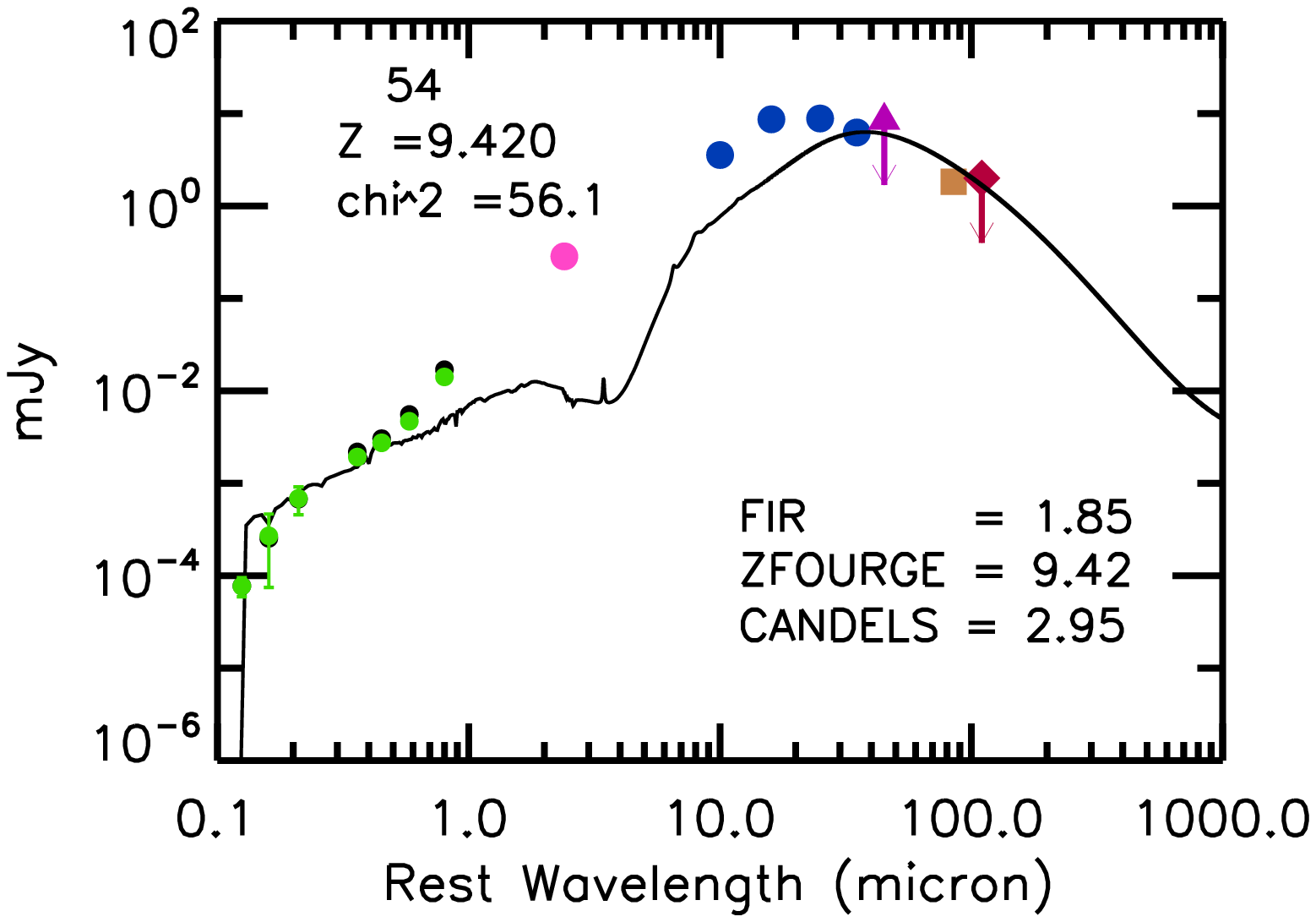} 
\includegraphics[width=2.8in,angle=0]{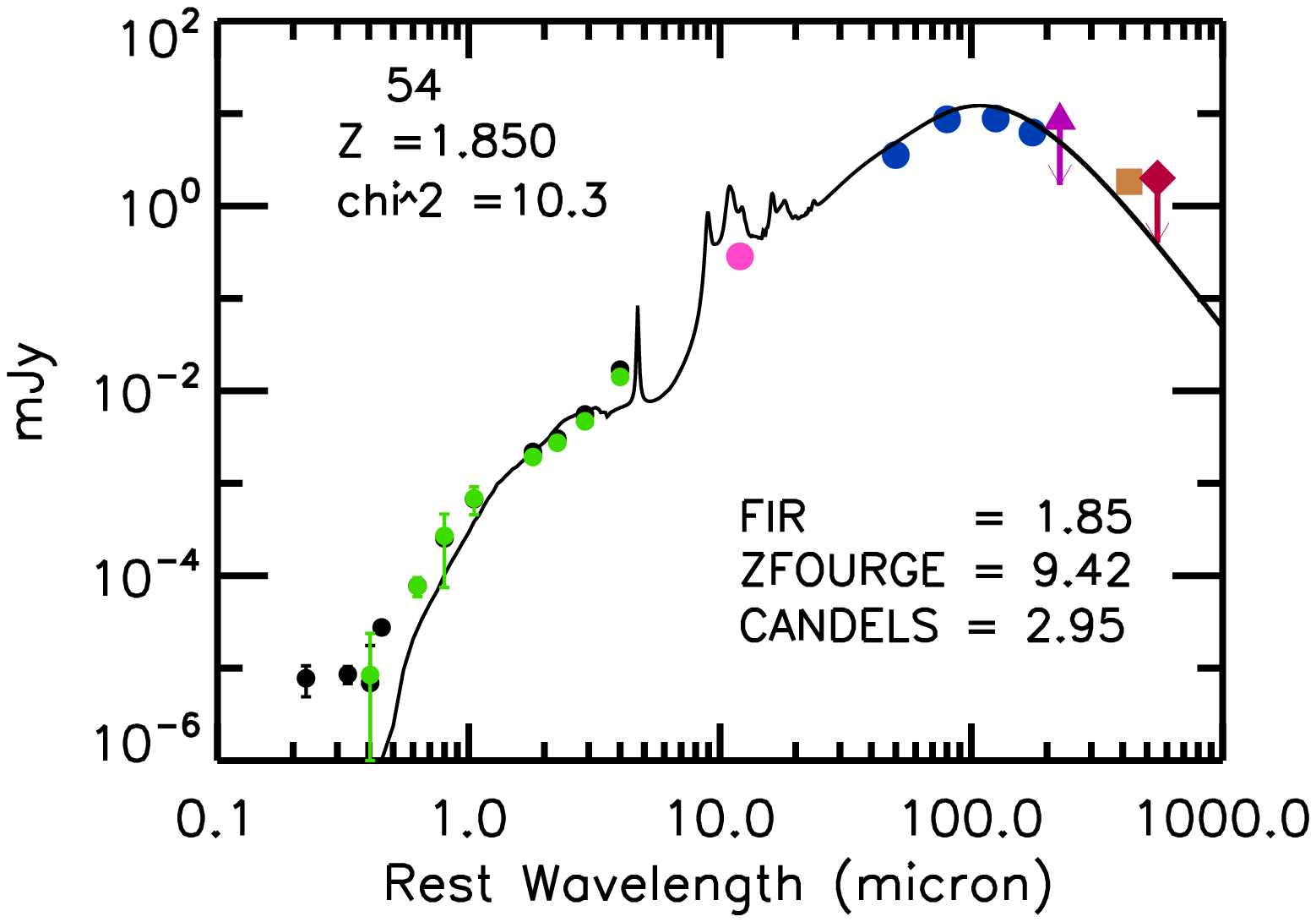} 
\includegraphics[width=2.8in,angle=0]{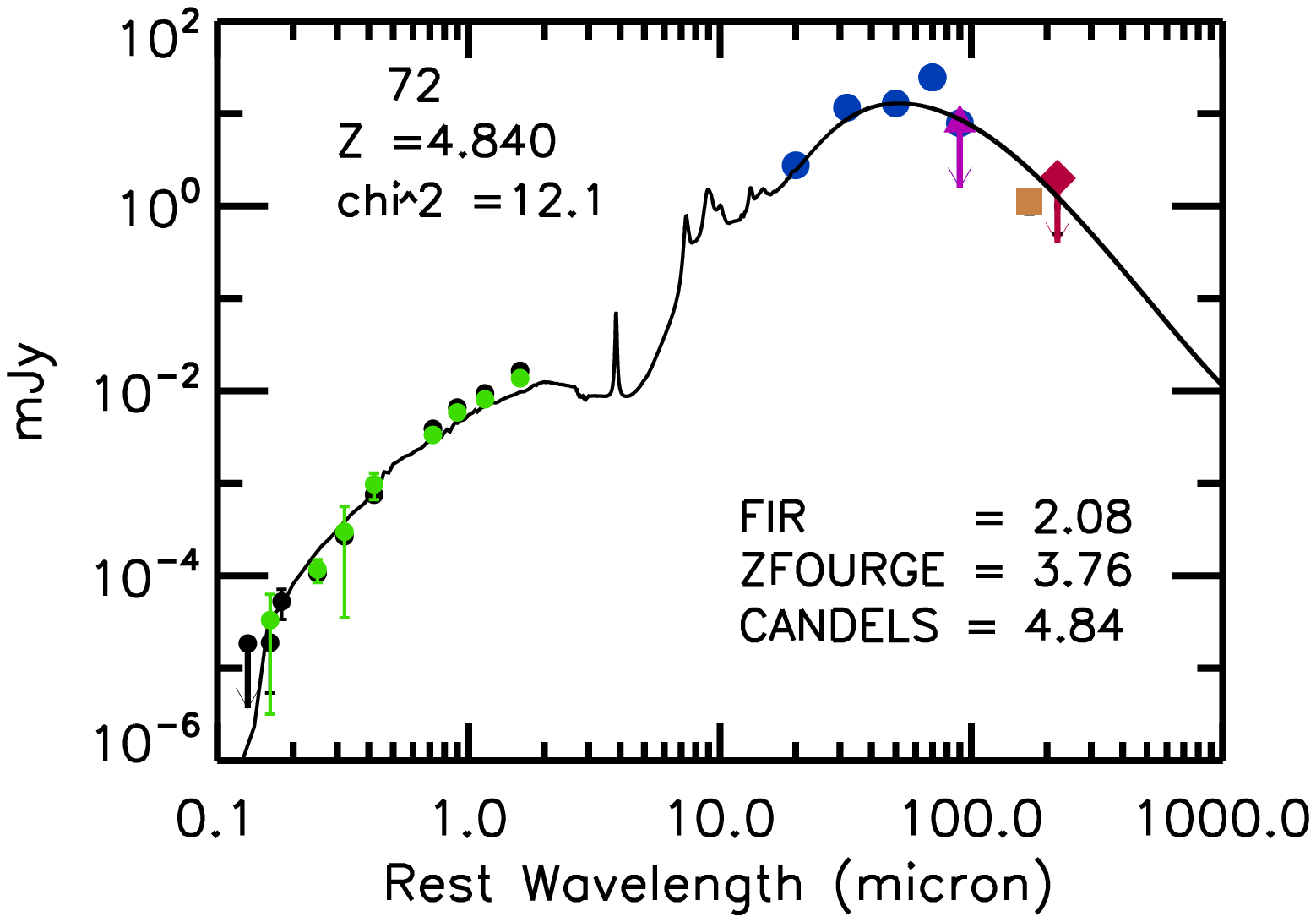} 
\hspace{3.8cm}\includegraphics[width=2.8in,angle=0]{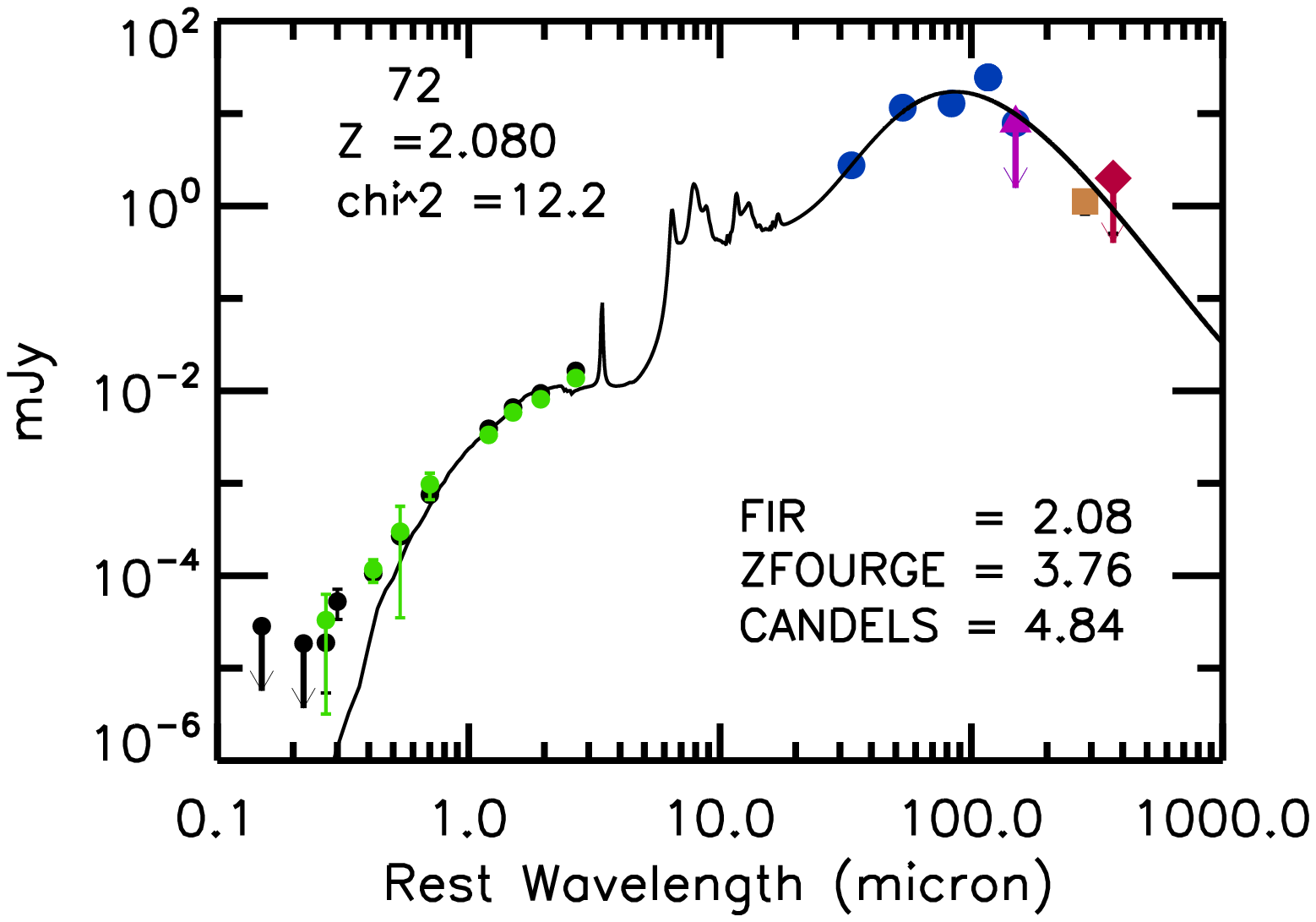} 
\caption{(Cont)
\label{wide_sedz2}
}
\end{figure*}

\begin{figure}[ht]
\centerline{\includegraphics[width=9cm,angle=0]{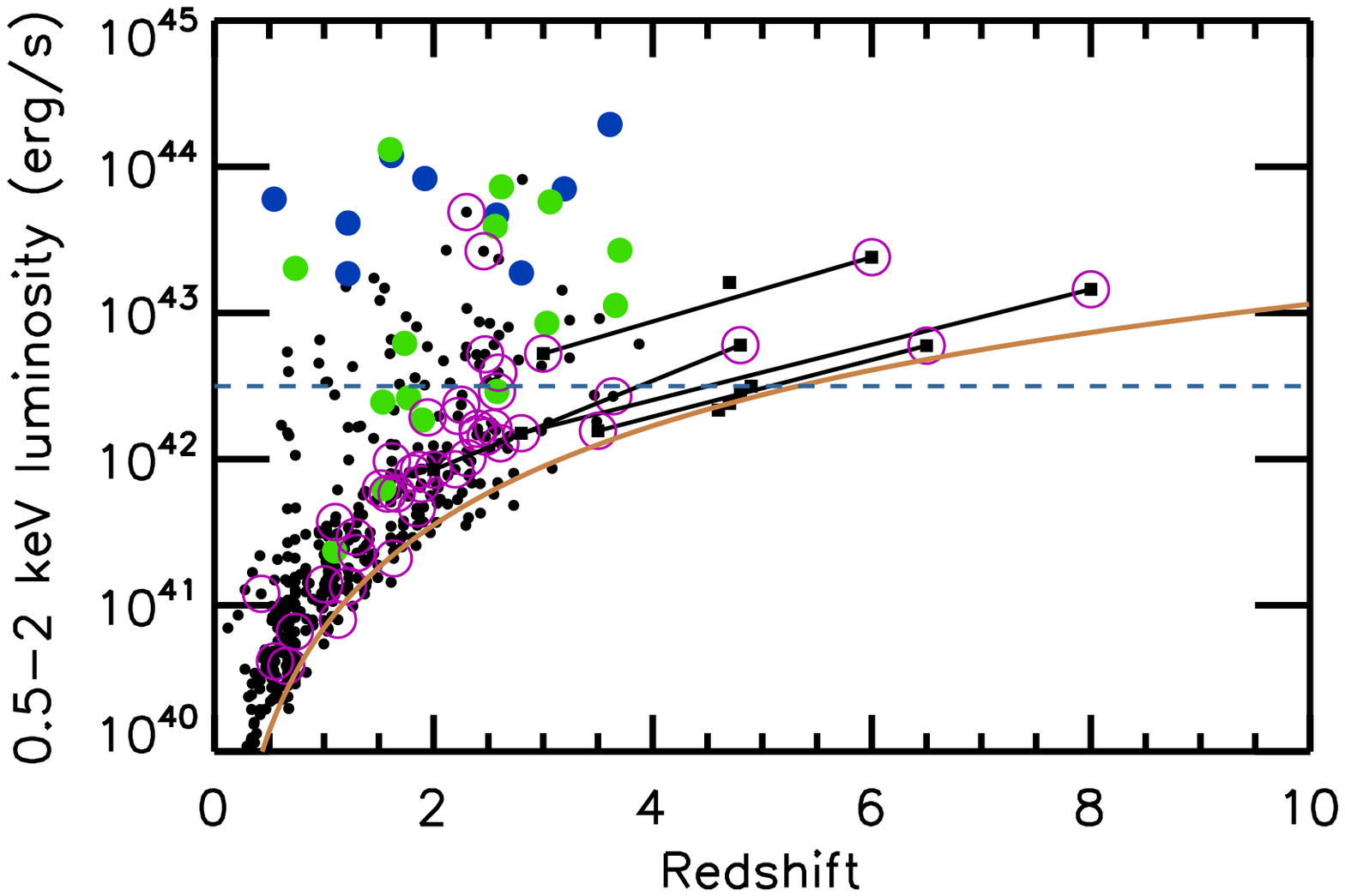}}
\centerline{\includegraphics[width=9cm,angle=0]{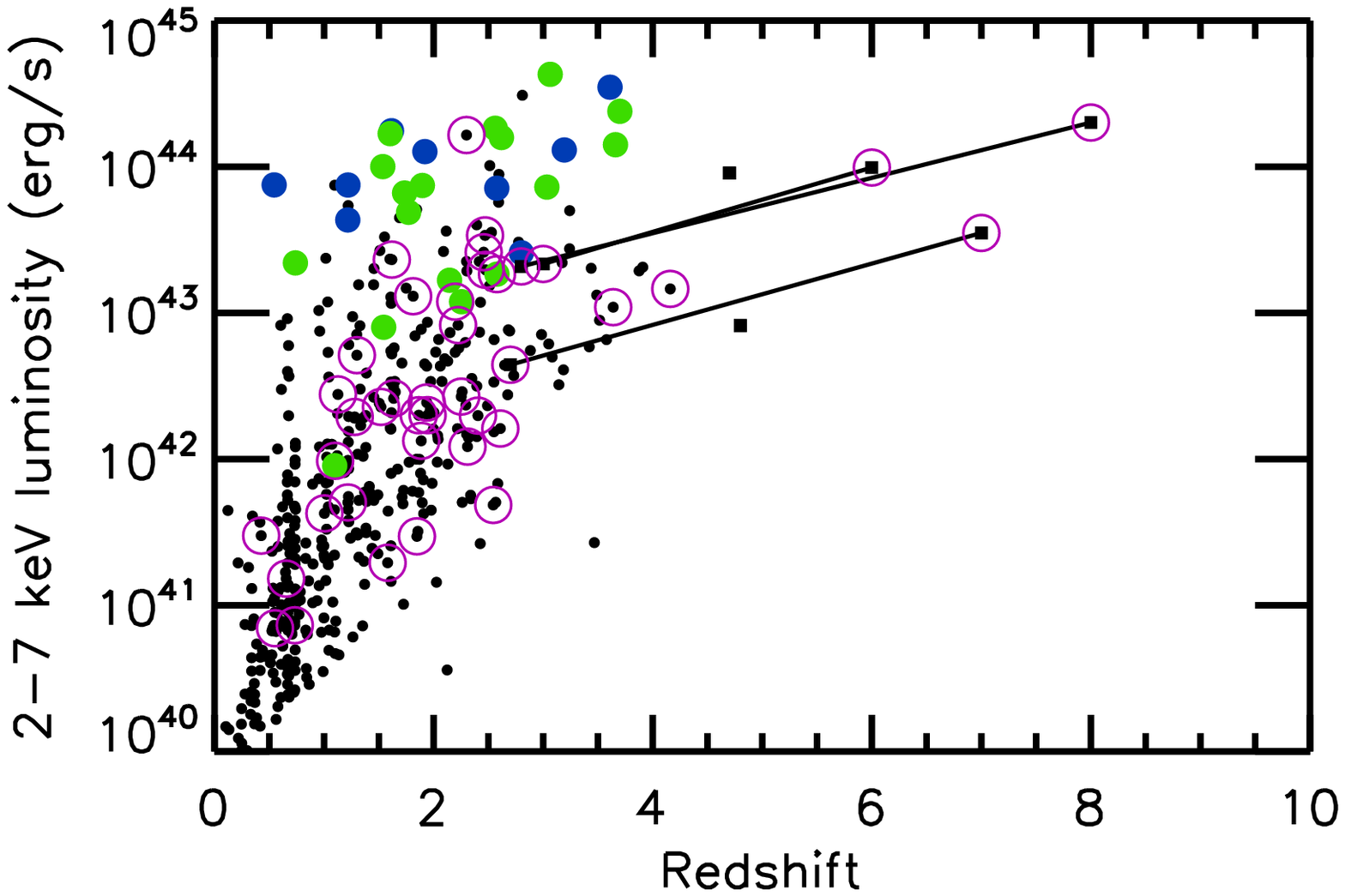}}
\caption{
(a) 0.5--2~keV and (b) 2--7~keV luminosity vs. redshift
(blue---broad-line AGNs; green---Seyfert type~2s; 
purple enclosing circles---ALMA sources). For the sources with
ambiguous redshifts, we show the endpoint redshifts
connected by a black line. For the 0.5--2~keV luminosities in (a),
the sample should be complete to the gold
curve, and we should see all sources $>10^{42.5}$~erg~s$^{-1}$
(blue dashed horizontal line) out to $z\sim6$, and all sources 
$>10^{43}$~erg~s$^{-1}$ out to $z\sim10$.
\label{z_ls}
}
\end{figure}

We first focus on the eight ALMA pre-selected X-ray detected $z>4.5$ candidates.
For these sources, we can construct the full SEDs from the optical 
to the FIR/submillimeter. These full SEDs, which we show in Figure~\ref{wide_sedz}
ordered by decreasing ALMA flux, can be used to eliminate 
some of the high-redshift photzs. The optical/NIR data, including all
four {\em Spitzer\/} IRAC bands, are taken preferentially
from the ZFOURGE (green circles) compilation of Straatman et al.\ (2016),
and otherwise from the CANDELS (black circles) compilation of Guo et al.\ (2013). 
The 24~$\mu$m and {\em Herschel\/} data (100 to 350~$\mu$m) are from 
Elbaz et al.\ (2011) using their 24~$\mu$m prior catalog, where possible, or, 
otherwise, a measurement made from the matched filter images at the ALMA position 
(blue circles). The 450~$\mu$m measurement from SCUBA-2 (purple triangle)
and the 850~$\mu$m measurement from ALMA (gold square) are both from C18.
The red diamonds show the AZTEC 1.1~mm measurements of Scott et al.\ (2010).
We fitted the SEDs with the  Bayesian energy-balance
MAGPHYS code of Da Cunha et al.\ (2008), fixing the 
redshift in the code independently to each of the ZFOURGE, CANDELS, and FIRz 
values for the source. In Figure~\ref{wide_sedz}, we show 
either the ZFOURGE or CANDELS-based fit in the left panels and the FIRz-based 
fit in the right panels (unless otherwise noted). 

The MAGPHYS code only includes star formation activity and  
not AGN activity, which could contribute to the optical/NIR/MIR portion 
of the SED. We return to this point below.
We use MAGPHYS, because it allows us to determine whether a good 
star formation fit can be found at the specified redshift
based on the $\chi^2$, and it also allows a visualization of departures from a 
simple star formation model.

The brightest ALMA source in the sample
is source~3 in Table~\ref{tab1} (C18~\#7
or L17~\#714). The $z_{\rm FIR}=3.37$ is consistent
with the ZFOURGE $z_{\rm phot}=3.48$; the CANDELS
$z_{\rm phot}=5.29$.
Fixing the redshift in MAGPHYS to the latter provides a reasonable
fit to all but the 24~$\mu$m flux, which is much too high
for a star-forming galaxy, yet the overall SED does not
appear to show AGN signatures. We adopt the ZFOURGE redshift 
of $z=3.48$, which produces an excellent fit to the full data.
We eliminate this source from the high-redshift candidate sample. 

The second brightest ALMA source is source~5 in Table~\ref{tab1}
(C18~\#17 or L17~\#657). Adopting $z_{\rm FIR}=7.99$ 
yields an observed 24~$\mu$m
flux that is too high compared to the model SED fit; this would only be possible if
the source were AGN dominated. 
The ZFOURGE $z_{\rm phot}=3.57$ does not provide a good fit to a 
star-forming galaxy---the MAGPHYS $\chi^2$ is
3.48---but a slight change in the redshift to 
$z=3.1$ produces an excellent fit. (There is a narrow range of $z=2.8$--3.2 
where there are good fits that reproduce the 24~$\mu$m flux.)  In this one case, 
we show the $z=3.1$ fit in the left panel of Figure~\ref{wide_sedz} rather than the 
ZFOURGE or CANDELS-based fit.  At the FIRz-based redshift of $z_{\rm FIR}=7.99$,
the star formation fit is poor, because of the 24~$\mu$m data point. 
However, the SED can be fit  by a mixed AGN and star-forming galaxy contribution. 
C18 show a fit to the source with a redshift of $z=7.8$ based on a mixed 
Type~2 QSO and a star-forming galaxy template (their Figure~34). 
This used the SWIRE Type~2 QSO and Arp~220 templates from Polletta et al.\ (2007).

Here we place the source in the redshift range $z=2.8-8$, where
the upper limit is based on the 68\% confidence
range in the FIRz fit. The source only becomes
visible at wavelengths longer than the CANDELS F125W 
band, consistent with this redshift range.

The third brightest ALMA source is not in Table~\ref{tab1}, since 
the only available optical/NIR $z_{\rm phot}<4.5$. This is C18~\#19
in Table~\ref{tab3} or L17~\#472. There is no CANDELS 
or H14 photz, while the ZFOURGE $z_{\rm phot}=4.47$ is 
just below the formal cutoff; meanwhile, the FIRz estimate is
$z_{\rm FIR}=6.69$.
There is no indication of AGN activity in the MIR
SED, and good fits to a star-forming galaxy
can be found over a wide range of redshifts. The source only
becomes visible above 3.6~$\mu$m, consistent with
it being at high redshift. We adopt a redshift range of $z=2.7$--7, 
where MAGPHYS gives $\chi^2<1$.

The fourth brightest ALMA source is not in Table~\ref{tab1}, since it is
not included in either the ZFOURGE or CANDELS catalogs.
This source is also not in the L17 catalog, but it corresponds to 
C18~\#44, who found $z_{\rm FIR}=5.26$. 
We have included the 3.6 and 4.5~$\mu$m fluxes from Ashby et al.\ (2013) and 
measured 1.25 and 1.6~$\mu$m fluxes from the CANDELS data in fitting the SED.  
Good MAGPHYS fits to star-forming galaxies can be found over a wide range of 
redshifts with $\chi^2<1$ from $z=3.5$--6.5.  For this source, we show the 
$z=3.5$ fit in the left panel and the $z=6.5$ fit in the right panel.

The fifth brightest ALMA source is source~2 in Table~\ref{tab1}
(C18~\#45 or L17~\#195). Here all optical/NIR estimates imply
$z_{\rm phot}>4.5$, but the FIRz estimate is $z_{\rm FIR}=3.09$. 
None of the MAGPHYS fits to star-forming galaxies are
good, and it seems likely the SED is a mixture of an AGN and
a star-forming galaxy with the AGN being required to explain the MIR 
excess for models that provide good fits to the rest-frame UV/optical data.
This is one of Pacucci et al.\ (2016)'s DCBH candidates. 
Pacucci et al.\ argue that the photz is likely higher than 
$z_{\rm phot}=5$, but Cappelluti et al.\ (2016) dismiss a high
photz estimate for the source as being a consequence of artifacts 
in the SED, without specifying the exact problem. However,
visual inspection of the deep {\em HST\/} images shows the 
source as being present in the F814W band, which would rule 
out redshifts above $z=6$. Here we place the source in the
redshift range $z=3$--6 where the lower redshift is a nominal 
choice placing the source in the low-redshift range.

\begin{deluxetable*}{ccccccccccc}
\renewcommand\baselinestretch{1.0}
\tablewidth{0pt}
\tablecaption{Final $z>4.5$ Candidates\label{tab4}}
\scriptsize
\tablehead{No. & L17 & C18 & R.A. & Decl. & $\log f_{\rm 0.5-2}$ & $\log f_{\rm 2-7}$ & $z$  & 850~$\mu$m  & $\log L_{\rm 0.5-2}$ & $\log L_{\rm 2-7}$ \\ & No. & No. & \multicolumn{2}{c}{(J2000)} & \multicolumn{2}{c}{(erg~cm$^{-2}$~s$^{-1}$)} & &  (mJy) & \multicolumn{2}{c}{(erg~s$^{-1}$)}  \\ (1) & (2) & (3) & (4) & (5) & (6) & (7) & (8) & (9) & (10) & (11)}
\startdata
1 & \nodata &  \nodata & 53.1971 & -27.8279 & -16.86 &  \nodata & 4.5(4.2--4.9) &   0.86$\pm$0.42 & 42.3 & $<42.9$ \cr
2 & \nodata  & \nodata & 53.1411 & -27.7644 & -16.86 &  \nodata & 4.6(4.0--4.9) &   0.97$\pm$0.35 &  42.3 & $<42.8$ \cr
3 & \nodata  &   \nodata & 53.1082 & -27.8251 & -16.83 &  \nodata & 4.7(4.6--4.8) &   0.15$\pm$0.26 & 42.3 & $<42.8$ \cr
4 & 490 &   \nodata &   53.1116 & -27.7678 & -16.01 &  -15.26 &  4.7(4.5--4.9) &  0.00$\pm$0.35 & 43.2 &   43.9\cr
5 & 527 &   72 &    53.1199 & -27.7430 & -16.46 &  \nodata & 2.0--4.8 &   1.11$\pm$0.29 & 41.9 -- 42.7 & $<42.6$ \cr
6 & 662 &    \nodata &   53.1479 & -27.8618 & -16.78 & -16.32 & 4.8(4.2--4.9) &  -0.65$\pm$0.34 & 42.4 & $42.9$ \cr
7 & \nodata  &   \nodata & 53.0876 & -27.7210 & -16.76 &  \nodata & 4.9(1.0--5.4) &   1.16$\pm$0.42 & 42.4 & $<43.1$ \cr
8 & \nodata  &  44 &   53.0872 & -27.8402 & -16.74 &  \nodata & 3.5--6.5 &   2.21$\pm$0.12 & 42.2 -- 42.7 & $<42.5$ -- $<43.6$ \cr
9 & 195 &    45 &    53.0410 & -27.8377 & -16.06 &  -15.45  &  3.0--6.0 &   2.43$\pm$0.21 & 42.7 -- 43.5 &   43.4 -- 44.0\cr
10 & 472 &   19 &   53.1088 & -27.8690 & \nodata &  -16.03 & 2.7--7.0 &   3.62$\pm$0.17 & $<41.5$ -- $<42.4$ &   42.7 -- 43.5\cr
11 & 657 &   17 &     53.1466 & -27.8710 & -16.54 &  -15.40 & 2.8--8.0 &   2.43$\pm$0.21 & 42.2 -- 43.2 &   43.3 -- 44.34\cr
\enddata
\tablecomments{
Columns: (1) Final $z>4.5$ candidate source number, 
(2) L17 X-ray catalog number, when available,
(3) C18 ALMA catalog number, when available,
(4) and (5) CANDELS R.A. and decl. from Guo et al.\ (2013), except for source~8, where they are the ALMA
R.A. and decl. from C18,
(6) and (7) logarithms of the 0.5--2~keV and 2--7~keV fluxes, if the source has a positive flux,
(8) adopted photz (if from ZFOURGE, then we include the 95\% confidence range in parentheses),
(9) ALMA 850~$\mu$m flux from C18, when available, or SCUBA-2 850~$\mu$m flux otherwise,
(10) and (11) logarithms of the 0.5--2~keV and 2--7~keV luminosities
calculated from Equations~1 and 2, respectively.
}
\end{deluxetable*}
\twocolumngrid 

The sixth brightest ALMA source is source~9 in Table~\ref{tab1}. 
It is not in the L17 catalog, but it corresponds to C18~\#52.
This source seems to be well described by a star-forming galaxy SED.
The higher photzs ($z_{\rm phot}=4.78$ from ZFOURGE and
$z_{\rm phot}=6.26$ from CANDELS) give a poor fit
at 24~$\mu$m, while $z_{\rm FIR}=3.63$ gives a good
overall fit.
While it might be possible to move this source to a higher redshift by 
including an AGN component, as was done for source~5 
(C18 \#17 or L17 \#657), the FIR shape 
appears more consistent with this being a star former.
We therefore drop this source from the $z>4.5$ candidate sample. 
 
The seventh brightest ALMA source is source~14 in Table~\ref{tab1} 
(C18~\#54 or L17~\#802). The SED is 
complex, and all of the MAGPHYS fits to star-forming galaxies are poor. 
It appears that this source has a strong AGN component.
However, the source is clearly detected in F606W,
which rules out $z>4.5$, and, in particular, the
ZFOURGE $z_{\rm phot}=9.4$. We therefore remove this source
from the high-redshift candidate sample.

The final ALMA detected source is source~4 of Table~\ref{tab1}
(L17~\#527). This source lies below
our submillimeter flux threshold and is not included
in Table~\ref{tab3} (it is source~72 in C18's Table~4). 
The SED is poorly fit at all redshifts and again may have AGN signatures. 
We adopt a redshift range of $z_{\rm phot}=2$ (H14) to
$z_{\rm phot}=4.8$ (CANDELS).

The remaining sources in Table~\ref{tab1} (excluding source~6 or L17~\#341) 
all lie at the low end of our high-redshift range ($z=4.5-5$) and have 
broadly consistent optical/NIR photzs. 
For these, we adopt the ZFOURGE photzs, 
but none of the discussion is dependent on this choice. 

In Table~\ref{tab4}, we summarize the 11 sources of our
final $z>4.5$ candidate sample. For sources where we adopt the ZFOURGE photzs,
we also provide the 95\% confidence range in the table.

In Figure~\ref{z_ls}, we plot X-ray luminosity versus redshift.
Above $z=4.5$, the figure is based on Table~\ref{tab4}, while
at lower redshifts, it is based on the redshift compilation
of Barger et al.\ (2019). Sources with
ambiguous redshifts are shown as endpoint redshifts connected
by a black line. The number density of luminous AGNs  
drops rapidly above $z=4$, and this is true even if we place the
ambiguous redshift sources at the high end redshift values.
The 0.5--2~keV sample should be complete to a flux
limit of $1.45\times10^{-17}$~erg~cm$^{-2}$~s$^{-1}$ throughout
the region (L17). Therefore, 
for the 0.5--2~keV luminosities in Figure~\ref{z_ls}(a), the sample 
should be complete to the gold curve, and we should see all sources 
above $10^{42.5}$~erg~s$^{-1}$ (blue dashed horizontal line) out to 
near $z=6$ and sources above $10^{43}$~erg~s$^{-1}$ out to $z=10$, 
so this is not a sensitivity issue.

Almost all of the sources in Table~\ref{tab4} lie just above $z=4.5$, 
and we are left with only five sources that might lie above $z=5$.
(Four based on the ranges discussed above, and one based on the 
ZFOURGE 95\% confidence limits.)  L17~\#657 and L17~\#195 have previously 
been considered as high-redshift candidates, with L17~\#195 considered as 
a possible candidate DCBH (e.g., Fiore et al.\ 2012; G15; Pacucci et al.\ 2016). 
The other three sources (L17~\#472, C18~\#44, and source~7 in Table~\ref{tab4}) are 
new candidates. Four sources are ALMA detected. They also have low
effective observed photon indices with the 2--7~keV flux being much 
stronger than the 0.5--2 keV flux. 
This would require them to have very strong X-ray obscuration, consistent with 
their red NIR colors and ALMA counterparts.

Many of the sources in Table~\ref{tab4} have 
$L_{\rm 0.5-2~keV}<10^{42.5}$~erg~s$^{-1}$ (Column~10),
where the X-ray luminosity could arise from X-ray
binary contributions (Barger et al.\ 2019). 
However, given the obscuration potential mentioned in the previous 
paragraph, the 2$\sigma$ upper limit on $L_{\rm 2-7~keV}$ could 
place them as weak AGNs. 
For our primary analysis, we will consider the seven sources with 
higher X-ray luminosities ($>10^{42.5}$~erg~s$^{-1}$) 
in one or other of these bands (i.e., sources 4, 5, 6, 8, 9, 10, and 11 in 
Table~\ref{tab4}). All but one of these are in the L17 catalog (Column~2 of 
Table~\ref{tab4}), and five of them are in the C18 ALMA catalog 
(Column~3 of Table~\ref{tab4}). Only five of them potentially
lie at $z>5$. However, we will also consider the 
effects of including the fainter sources on our conclusions.

We now consider the constraints imposed by these sources
on the high-redshift AGN luminosity density and on the
obscuration in high-redshift candidates.

\section{Discussion}
\label{secdisc}
As can be seen from Figure~\ref{z_ls} and Table~\ref{tab4}, there 
are very few X-ray luminous AGNs at $z>4.5$.
Only L17~\#195 (C18~\#45) and L17~\#657 (C18~\#17),
if they are placed at the highest possible redshifts,
could have quasar luminosities ($L_{\rm 2-7~keV}> 10^{44}$~erg~s$^{-1}$).
This is consistent with previous findings
(e.g., Barger et al.\ 2003; Fiore et al.\ 2012; Vito et al.\ 2018)
that the number density of luminous AGNs drops rapidly above $z=3$;
by $z=5$, it has fallen by around an order of magnitude.

\begin{figure}[ht]
\centerline{\includegraphics[width=9cm,angle=0]{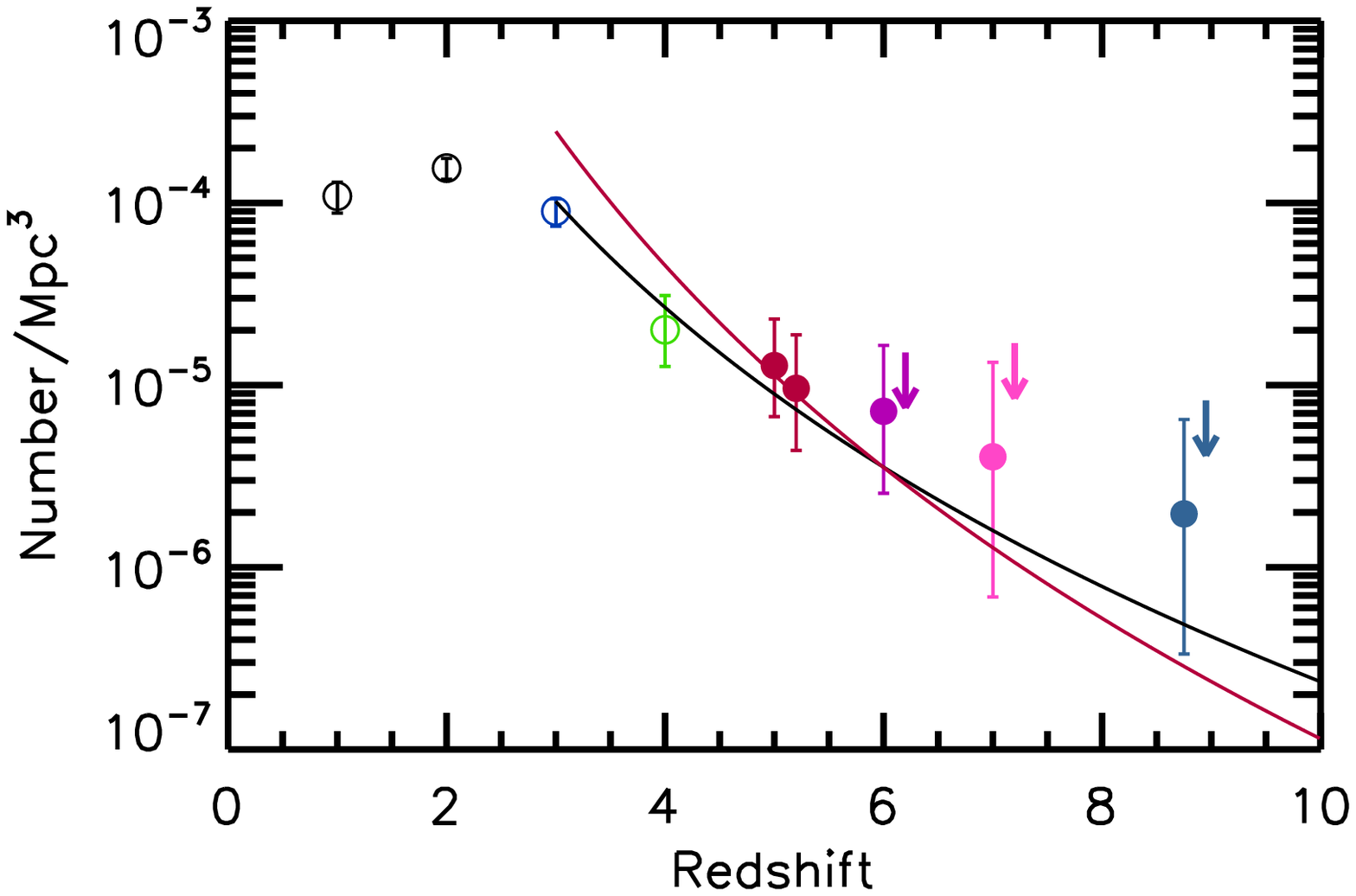}}
\caption{
Evolution of the total number of sources with a $>2\sigma$ detection 
in the band and 
$\log L_{\rm 0.5-2~keV}>42.5$~erg~s$^{-1}$
or $\log L_{\rm 2-7~keV}>42.5$~erg~s$^{-1}$
vs. redshift in the following
redshift intervals: 0.5--1.5 (black), 1.5--2.5 (black), 2.5--3.5 (blue),
3.5--4.5 (green), 4.5--5.5 (red), 5.5--6.5 (purple),
6.5--7.5 (pink), and 7.5--10 (navy blue). The evolution 
from $z=3$--6 can be fitted with the power law 
$0.9\times10^{-4} ((1+z)/4)^{-6}$~Mpc$^{-3}$ (black curve). 
The red curve shows the Vito et al. (2018) slope of
-7.59.
Solid circles show the number density based on Table~\ref{tab4}, 
while the open circles at lower redshifts are based on the L17 
catalog using the redshifts given in Barger et al.\ (2019). For the 
high-redshift intervals, we show the number densities both where
all the sources lie at the higher redshift end of their range and (displaced)
where all the sources lie at the lower redshift end of their range (for
these, we plot 95\% confidence upper limits; downward pointing arrows).
\label{highz_evol}
}
\end{figure}

This is most easily seen in Figure~\ref{highz_evol},
where we show the evolution of the total number density of sources with
a $>2\sigma$ detection in the band and
$\log L_{\rm 0.5-2~keV}>42.5$~erg~s$^{-1}$ 
or $\log L_{\rm 2-7~keV}>42.5$~erg~s$^{-1}$,
with the comoving volume computed for an area of 102~arcmin$^2$.
This lower luminosity limit was chosen to separate cleanly AGNs from
star-forming galaxies. The number density of AGNs
falls as $0.9\times10^{-4} ((1+z)/4)^{-6}$~Mpc$^{-3}$ 
between $z=3$ and $z=6$ (black curve).
(Vito et al.\ 2018 obtained a slightly steeper slope of $-7.59$ for this relation, 
which we show as the red curve in Figure~\ref{highz_evol})
Including all of the lower luminosity candidates under the assumption that they 
are heavily obscured would increase the $z=5$ point by a factor of 1.75, which 
would produce a slower evolution at $z<5$ and a more abrupt break to the 
lower values at the higher redshifts.

Even including all 5 galaxies that might be at $z>5$
in the $z=$5--7 redshift interval gives only a small
contribution to the required photoionization at these redshifts.
Computing the ionization rate following the assumptions about
the AGN SED shape and ionizing escape fraction ($f=1$) used in 
Giallongo et al.\ (2019), we find a photoionization rate of 
$1.7 \times 10^{-14}$~s$^{-1}$.
This extreme upper limit is only about 10\%
of the value required by the IGM ionization level 
(e.g., Wyithe \& Bolton 2011, who find $1.8 \times10^{-13}$~s$^{-1}$
at $z=6$), emphasizing that such faint high-redshift AGNs do not make
a significant contribution to the ionization.

\begin{figure}[ht]
\centerline{\includegraphics[width=9cm,angle=0]{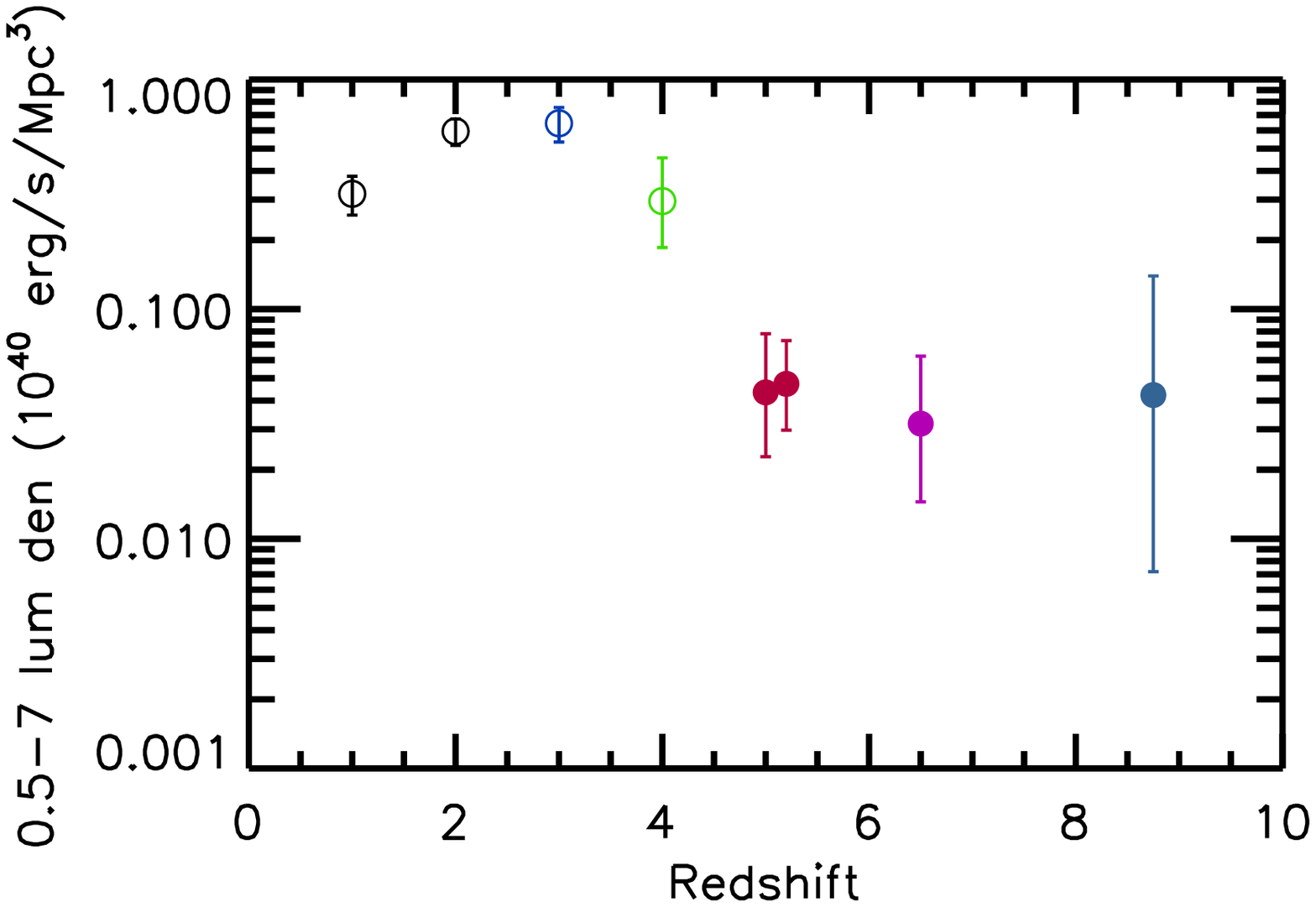}}
\caption{
Evolution of the 0.5--7~keV luminosity density of sources with
$\log L_{\rm 0.5-2~{\rm keV}}>42.5$~erg~s$^{-1}$
or $\log L_{\rm 2-7~{\rm keV}}>42.5$~erg~s$^{-1}$
vs. redshift in the following
redshift intervals: 0.5--1.5 (black), 1.5--2.5 (black), 
2.5--3.5 (blue), 3.5--4.5 (green), 4.5--5.5 (red), 
5.5--7.5 (purple), and 7.5--10 (navy blue). Solid circles
show the luminosity density based on the high-redshift values
in Table~\ref{tab4}, while the open
circles at lower redshifts are based on the L17 catalog
using the redshifts given in Barger et al.\ (2019). The errors are 68\%
confidence limits based on the number of sources in the
redshift interval. The displaced red point in the $z=4.5$--5.5 interval 
shows the value when lower luminosity sources are included.
\label{lumden_evol}
}
\end{figure}

There are only a small number of possible sources above $z=5$
(between zero and five), so the uncertainties are large.
For the highest redshift intervals, the $2\sigma$ upper limits
obtained (downward pointing arrows) if we assume all the sources 
are at the lower redshift end of their range are not substantially different
than the measured values obtained if we assume all the sources are at
the higher redshift end of their range. The power law fit to the $z=3$--6 points 
is consistent with the $z>5$ points, regardless of whether we place 
the sources at the high redshift or low redshift end of their ranges.
The mean density of AGNs in the $z=5$--10 interval is 
$4.8\times10^{-6}$~Mpc$^{-3}$ with a 68\% confidence range of 
2.7--$8.1\times10^{-6}$~Mpc$^{-3}$, if we place
all the sources at their highest possible redshifts, compared with a 95\%
confidence upper limit of $3.5\times10^{-6}$~Mpc$^{-3}$, 
if we place them all at their lowest possible redshifts.

The X-ray luminosity density also drops rapidly. In Figure~\ref{lumden_evol},
we show the evolution of the 0.5--7~keV luminosity density, which drops
by a factor of 25 or more between its peak value near $z=2$--3 to the
values at $z>5$. In the $z=5$--10 range, we find a value from 0 to
$3.8\times10^{38}$~erg~s$^{-1}$~Mpc$^{-3}$, depending on whether we
place the sources at low or high redshift.

\begin{figure}
\centerline{\includegraphics[width=3.8in,angle=0]{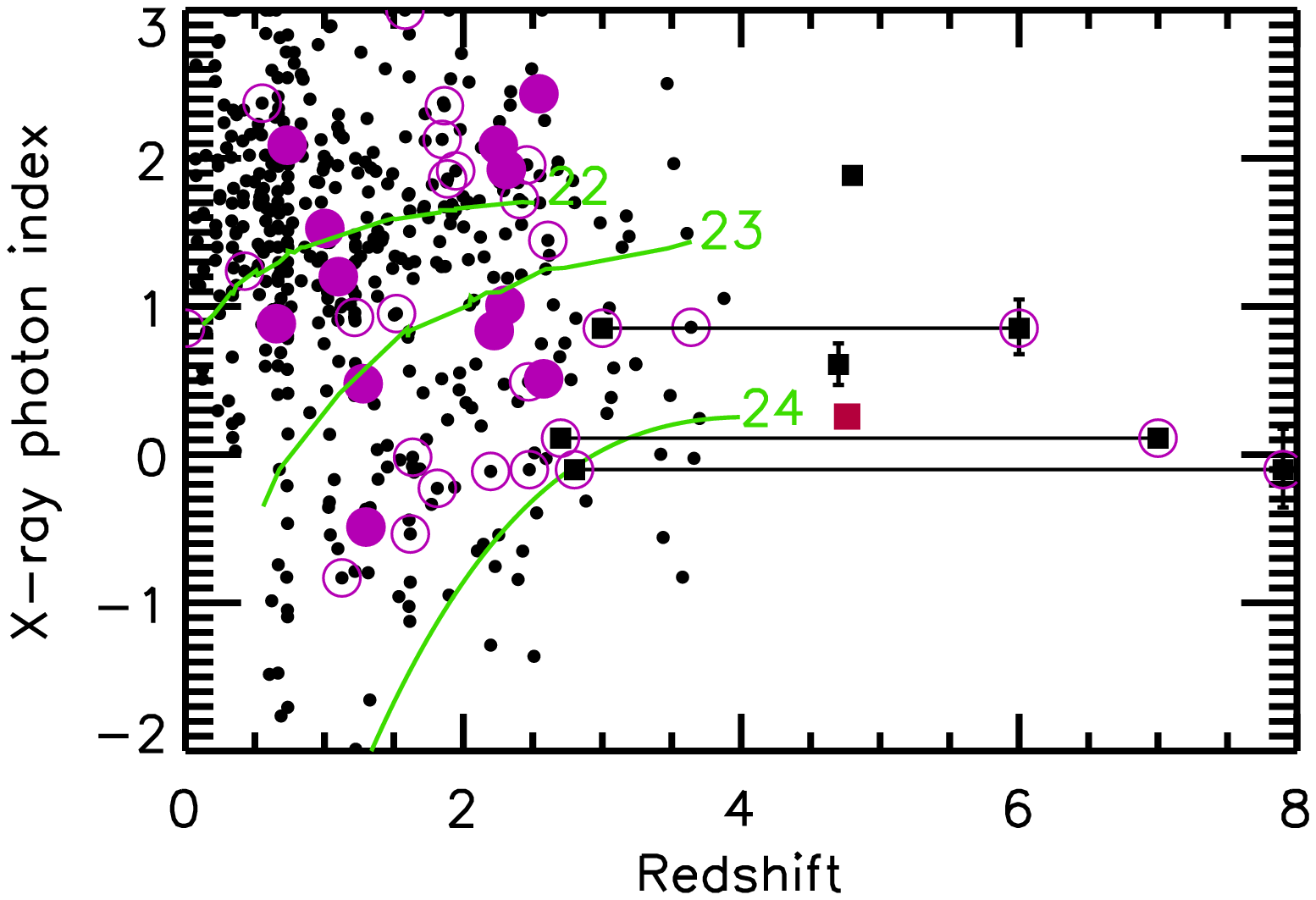}}
\caption{Distribution of effective observed photon index from L17
vs. redshift for the central region X-ray sources detected in the 2--7~keV band. 
Below $z=4.5$, we use the redshifts from Barger et al.\ (2019) based
on the speczs, then H14 photzs, and finally ZFOURGE photzs
(black small circles).
Above $z=4.5$, we use Table~\ref{tab4} (black large squares), 
and when there is
a range of redshifts, we show the end points connected by a black line.
For the $z>4.5$ sources that are detected in both the 0.5--2 and 2--7 keV 
bands, we show the $\pm1\sigma$ error bars. For the remaining $z>4.5$
sources, we show the estimated indices from L17.
Purple circles denote ALMA detected sources,
with the solid symbols showing those with a specz. The red square shows 
the high-redshift Compton-thick source of Gilli et al.\ (2011, 2014).
The green curves from top to bottom show absorption of
$\log N_H=22, 23,$ and 24~cm$^{-2}$
computed from the effective observed photon index assuming an intrinsic 
power law of $1.8$.
\label{xray_gamma}
}
\end{figure}

In Figure~\ref{xray_gamma}, we plot the effective observed photon 
index from L17 versus redshift for the candidate $z>4.5$ sources in 
Table~\ref{tab4} that are detected in the 2--7~keV band 
(black large squares). L17 calculated the effective observed photon
indices and errors from the hard to soft band ratios, assuming the 
0.5--7~keV spectra of the X-ray sources are 
power laws modified by only Galactic absorption. 
For sources detected in only one band, they made
a best guess estimate based on the mode values.
For the sources where we are using the 
ZFOURGE photzs, we show only a single point. For the remaining sources
where there is a range of redshifts, we show the
end points connected by a black line.
However, regardless of their redshift placement, these sources are unusual
compared to the lower redshift AGNs. We illustrate this in 
Figure~\ref{xray_gamma} by also showing the central region 
2--7~keV sample at lower redshifts (black small circles). We include
green curves on the plot to denote absorption of 
$\log N_H = 22$, 23, and 24~cm$^{-2}$ 
(top to bottom), which we computed
from the effective observed photon index assuming an intrinsic power law of $1.8$.
Three of the five sources are ALMA sources (purple enclosing circles), which 
implies high star formation rates. However,
two of these are also Compton-thick (L17~\#472 and L17~\#657).
The third (Luo~\#195) has an $N_H$ value between $10^{23}$ and 10$^{24}$~cm$^{-2}$,
which is comparable to most of the lower redshift ALMA detected hard X-ray sources 
(C18). For comparison, we also plot the high-redshift Compton-thick source of 
Gilli et al.\ (2011, 2014; red square).
In contrast, neither of the two $z<5$ sources detected in the 2-7~keV band are 
ALMA detected, and one (L17~\#662) has a low effective observed photon index 
consistent with it having little absorption.

\begin{figure}
\centerline{\includegraphics[width=3.8in,angle=0]{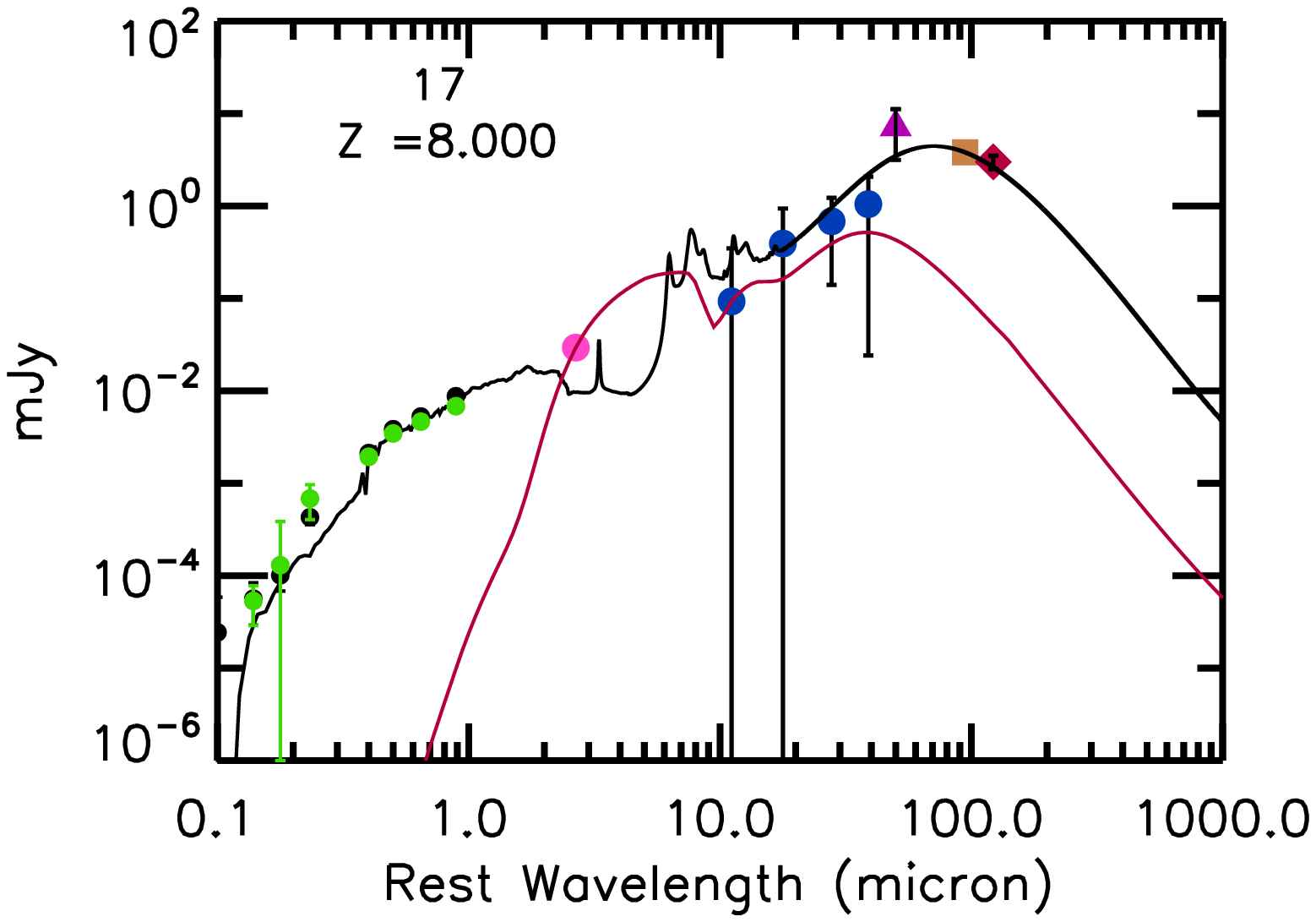}}
\centerline{\includegraphics[width=3.8in,angle=0]{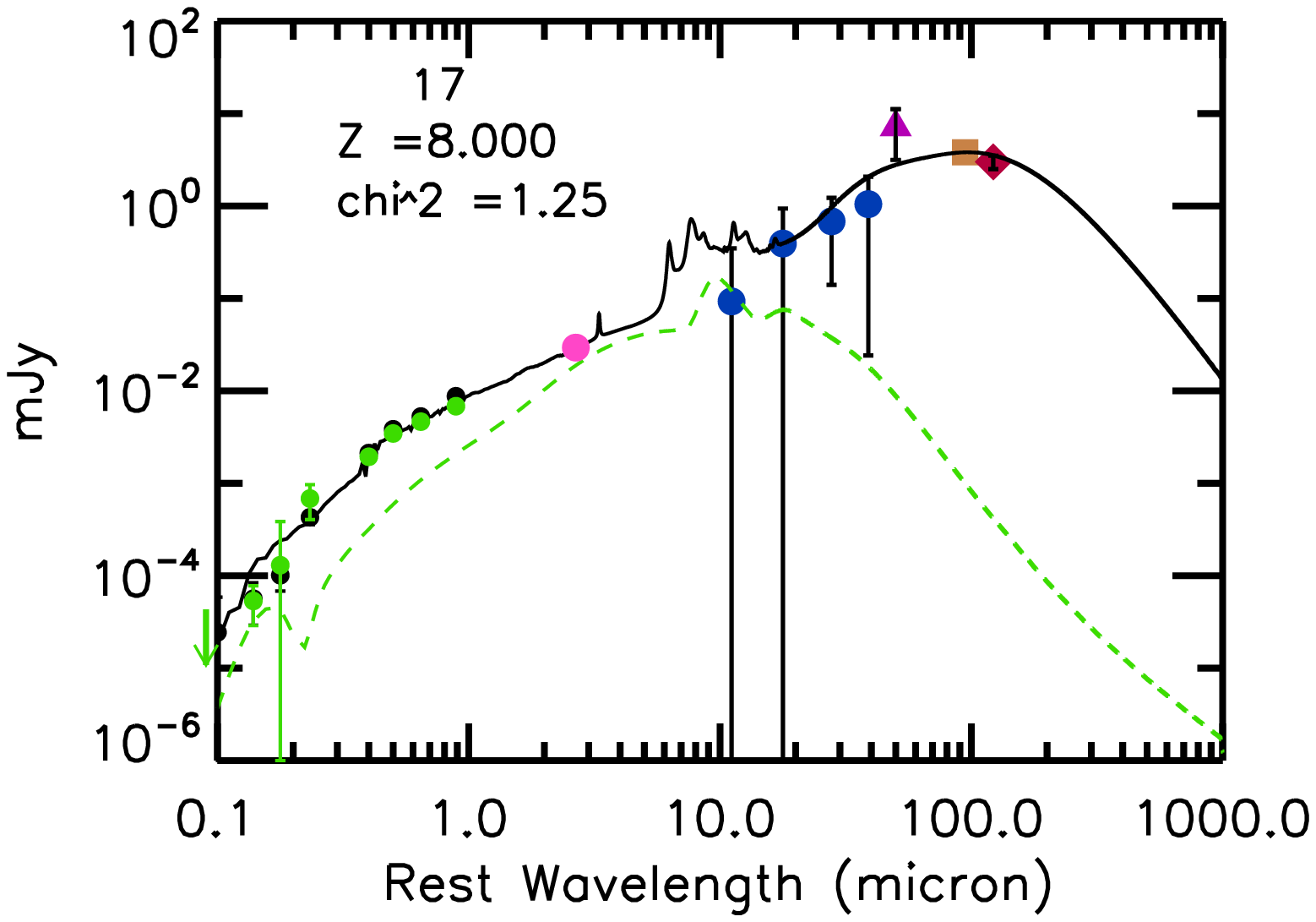}}
\caption{Maximum torus fit to L17~\#657 (C18~\#17) 
placed at $z=8$. (a) The red curve shows the maximum 
bolometric luminosity torus from Fritz et al.\ (2006) that can be included 
at this redshift. The black curve shows the best fit MAGPHYS star-forming 
galaxy SED. (b) The black curve shows a simultaneous 
star formation and torus fit made using the SED3FIT program of
Berta et al.\ (2013). The dashed green curve shows the torus component
alone.
\label{torus}
}
\end{figure}

\begin{figure}
\centerline{\includegraphics[width=3.8in,angle=0]{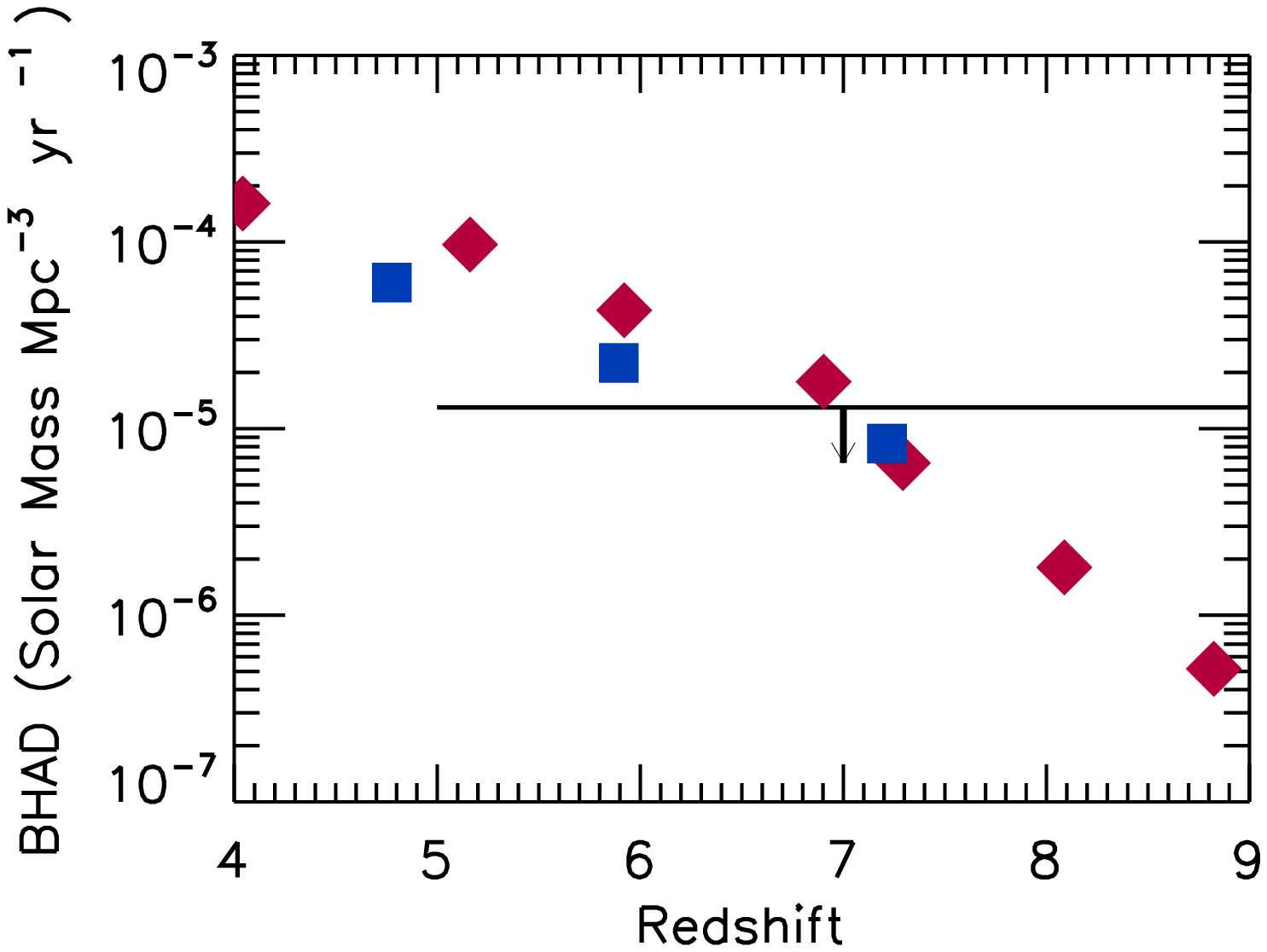}}
\caption{Upper bound to the black hole accretion density in the interval $z=5$--8
(black line with downward pointing arrow) compared with predictions from 
model simulations (blue squares---Sijacki et al.\ 2015; 
red diamonds---Volonteri et al.\ 2016).
The upper bound is based on including all possible sources in 
the high-redshift interval and using the maximum bolometric correction. 
It should be an extreme upper limit, unless there are significant numbers 
of extreme Compton-thick sources.
\label{bhad_fig}
}
\end{figure}

In order to compare with models, we need to convert the measured X-ray luminosity
density to the black hole accretion density, which is the predicted quantity. 
Given the complexity of the sources and our desire to have as
robust an upper bound on the black hole accretion density as possible, 
we do not attempt to apply
simple bolometric corrections (e.g., the Brightman et al.\ 2017 values 
calculated for local Compton-thick AGNs)
to calculate the bolometric AGN luminosity from the X-ray luminosity.
This also avoids us having to apply uncertain absorption corrections 
to the X-ray luminosities.
We also note that using the MIR to X-ray conversions of
Gandhi et al.\ (2009) and Asmus et al.\ (2015) would produce bolometric
luminosity density estimates several orders of magnitude lower
than the maximal torus estimate given below. 

Instead, we fit torus models 
to the observed SEDs of the sources to determine the maximum possible
AGN bolometric luminosity that can be included. 
We use the torus templates of Fritz et al.\ (2006). 
We fit the torus to the observed SEDs without including the 
star formation contributions from the galaxy. 
As we illustrate in Figure~\ref{torus}, including those contributions
could substantially reduce the limits on the torus.
Our torus fitted AGN bolometric luminosities are strictly an upper bound 
constrained primarily by the 24~$\mu$m and shorter
wavelength {\em Herschel\/} fluxes. Combining the AGN bolometric 
luminosities for the five sources, 
we find a maximum bolometric luminosity density of 
$7.7\times10^{40}$~erg~s$^{-1}$~Mpc$^{-3}$ in the interval $z=5$--10.
This is an extreme upper limit, 
because we have included all possible redshift sources at the 
high-redshift end, and because of our maximal bolometric correction.
However, it could still omit extreme Compton-thick AGNs.

We have also made simultaneous fits to the data using the Berta
et al. (2013) SED3FIT program, which includes both the MAGPHYS star formation
models and the Fritz et al.\ (2006) torus models. A comparison with the maximum
torus model for source L17~\#657 (C18~\#17) is shown in Figure~\ref{torus}.
In general, the combined star formation and torus fits make a small decrease
of just over a factor of 2 in the torus luminosity density 
($3.5\times10^{40}$~erg~s$^{-1}$~Mpc$^{-3}$) relative to the maximum torus fits.

If we assume a radiative efficiency of $\epsilon=0.1$, then  the
maximum torus luminosity density converts
to a black hole accretion density of 
$1.3\times10^{-5}$~M$_\sun$~yr$^{-1}$~Mpc$^{-3}$ in the interval $z=5$--10. 
We do not attempt to assign a statistical error to this, since the primary 
uncertainties lie in the photzs and in the use of the maximal torus luminosities.
Even with this extreme upper bound, which is considerably
higher than previous estimates, such as those of 
Vito et al.\ (2016), who give $8.16\times10^{-7}$~M$_\sun$~yr$^{-1}$~Mpc$^{-3}$
in the interval $z = 4.5$--5.5 and $1.76\times 10^{-7}$~M$_\sun$~yr$^{-1}$~Mpc$^{-3}$
in the interval $z = 5.5$--6.5, this is lower than the 
predicted values at $z=5$ from simulations (e.g., Sijacki et al.\ 2015; Volonteri et al.\ 2016).
We illustrate this in Figure~\ref{bhad_fig}.

\section{Summary}
\label{secsummary}
\begin{itemize}

\item[$\bullet$] 
We searched for high-redshift ($z>4.5$) X-ray AGNs 
in the CDF-S by analyzing direct X-ray detections
and by probing more deeply using samples pre-selected at other
wavelengths (the NIR/MIR and the submillimeter).

\item[$\bullet$] 
We conducted a detailed review using the SEDs from the optical to 
the FIR/submillimeter to determine if previous photzs were plausible.

\item[$\bullet$] 
Using our final candidate high-redshift sample, we found that the number
density of sources falls rapidly between $z=3$ and $z=6$.

\item[$\bullet$] 
The three highest redshift candidates are both submillimeter 
sources and highly obscured X-ray sources.  Two of the sources,
if they indeed lie at such high redshifts, must be Compton-thick sources, 
analogous to the $z=4.762$ source found by Gilli et al.\ (2011, 2014).

\item[$\bullet$] 
The measured X-ray light density in the interval $z=5$--8
implies a very low black hole accretion density with very little growth
in the black hole mass density in this redshift range.

\end{itemize}

\vskip 5cm
\acknowledgements
We would like to thank the referee for a very useful and interesting
report that helped us to improve the manuscript.
We gratefully acknowledge support from NASA grant NNX17AF45G (L.~L.~C.),
NSF grant AST-1313150 (A.~J.~B.),
CONICYT grant Basal-CATA PFB-06/2007 (F.~E.~B, J.~G.-L.), 
Programa de Astronomia FONDO ALMA 2016 31160033 (J.~G-L),
and the Ministry of Economy, Development, and Tourism's Millennium Science 
Initiative through grant IC120009, awarded to The Millennium Institute of Astrophysics, 
MAS (F.~E.~B.).
Este trabajo cont{\'o} con el apoyo de CONICYT + Programa de Astronom{\'i}a + 
Fondo CHINA-CONICYT CAS16026 (J.~G.-L.).
Support for this research was also provided by the University of Wisconsin-Madison,
Office of the Vice Chancellor for Research and Graduate Education with funding
from the Wisconsin Alumni Research Foundation,
the John Simon Memorial Guggenheim Foundation, and the Trustees
of the William F. Vilas Estate.
A.~J.~B would like to thank the Lorentz Center for
the stimulating workshop ``Monsters of the Universe:  The Most Extreme Star Factories''
that benefited this work.
ALMA is a partnership of ESO (representing its member states), 
NSF (USA) and NINS (Japan), together with NRC (Canada), 
MOST and ASIAA (Taiwan), and KASI (Republic of Korea), in cooperation 
with the Republic of Chile. The Joint ALMA Observatory is operated by 
ESO, AUI/NRAO and NAOJ.
The James Clerk Maxwell Telescope is operated by the East Asian Observatory 
on behalf of The National Astronomical Observatory of Japan, Academia Sinica 
Institute of Astronomy and Astrophysics, the Korea Astronomy and Space 
Science Institute, the National Astronomical Observatories of China and the 
Chinese Academy of Sciences (Grant No. XDB09000000), with additional funding 
support from the Science and Technology Facilities Council of the United Kingdom 
and participating universities in the United Kingdom and Canada.
The W.~M.~Keck Observatory is operated as a scientific
partnership among the California Institute of Technology, the University
of California, and NASA, and was made possible by the generous financial
support of the W.~M.~Keck Foundation.
The authors wish to recognize and acknowledge the very significant 
cultural role and reverence that the summit of Maunakea has always 
had within the indigenous Hawaiian community. We are most fortunate 
to have the opportunity to conduct observations from this mountain.



\begin{references}

\reference{asmus15}
Asmus, D., Gandhi, P., Honig, S. F.,  et al.\ 2015, \mnras, 454, 766

\reference{ashby13}
Ashby, M. L. N., Willner, S. P., Fazio, G. G., et al.\ 2013, \apj, 769, 80

\reference{banados18}
Ba\~nados, E., Connor, T., Stern, D., et al.\ 2018, \apj, 856, L25

\reference{banados16}
Ba\~nados, E., Venemans, B. P., Decarli, R., et al.\ 2016, \apjs, 227, 11

\reference{barger19}
Barger, A. J., Cowie, L. L., Bauer, F. E., \& Gonz{\'a}lez-L{\'o}pez, J.\ 2019, \apj, 887, 23

\reference{barger02}
Barger, A. J., Cowie, L. L., Brandt, W. N., et al.\ 2002, \aj, 124, 1839

\reference{barger03}
Barger, A. J., Cowie, L. L., Capak, P., et al.\ 2003, \apj, 584, L61

\reference{barger17}
Barger, A. J., Cowie, L. L., Owen, F., Hsu, L.-Y., \& Wang, W.-H.\ 2017,
\apj, 835, 95 

\reference{berta13}
Berta, S., Lutz, D., Santini, P., et al.\ 2013, A\&A, 551, 100

\reference{brandt15}
Brandt, W. N., \& Alexander, D. M.\ 2015, ARA\&A, 23, 1

\reference{bright17}
Brightman, M., Balakovic, M., Ballantyne, D. R., et al.\ 2017, \apj, 844, 10

\reference{bromm03}
Bromm, V., \& Loeb, A.\ 2003, \apj, 596, 34

\reference{capak11}
Capak, P. L., Riechers, D., Scoville, N. Z., et al. 2011,\ Nature, 470, 233

\reference{cap16}
Cappelluti, N., Comastri, A., Fontana, A., et al.\ 2016, \apj, 823, 95

\reference{circ19}
Circosta, C., Vignali, C., Gilli, R., et al.\ 2019, A\&A, 623, 172

\reference{cowie12}
Cowie, L. L., Barger, A. J., \& Hasinger, G.\ 2012, \apj, 748, 50

\reference{cowie17}
Cowie, L. L., Barger, A. J., Hsu, L.-Y., et al.\ 2017, \apj, 837, 139

\reference{cowie18}
Cowie, L. L., Gonz{\'a}lez-L{\'o}pez, J., Barger, A. J., et al.\ 2018, \apj, 865, 106 (C18)

\reference{dacun2008}
Da Cunha, E., Charlot, S., \& Elbaz, D.\ 2008, \mnras, 388, 1595

\reference{dahlen13}
Dahlen, T., Mobasher, B., Faber, S. M., et al.\ 2013, \apj, 775, 93

\reference{dickinson03}
Dickinson, M., Giavalisco, M., \& GOODS Team 2003, in The
Mass of Galaxies at Low and High Redshift, ed. R. Bender \&
A. Renzini (Berlin: Springer), 324

\reference{elbaz11}
Elbaz, D., Dickinson, M., Hwang, H. S., et al.\ 2011, A\&A, 533, A119

\reference{fiore12}
Fiore, F., Puccetti, S., Grazian, A., et al.\ 2012, A\&A, 537, A16

\reference{fritz06}
Fritz, J., Franceschini, A., Hatziminaoglou, E.\ 2006, \mnras, 366, 767

\reference{gandhi09}
Gandhi P., Horst H., Smette A., et al.\ 2009
A\&A, 502, 457

\reference{giallongo15}
Giallongo, E., Grazian, A., Fiore, F., et al.\ 2015, A\&A, 578, A83 (G15)

\reference{giallongo19}
Giallongo, E., Grazian, A., Fiore, F., et al.\ 2019, \apj, 884, 19

\reference{giavalisco04}
Giavalisco, M., Dickinson, M., Ferguson, H. C., et al.\ 2004, \apj, 600, L93

\reference{gilli11}
Gilli, R., Su, J., Norman, C., et al.\ 2011, \apj, 730, L28

\reference{gilli14}
Gilli, R., Norman, C., Vignali, C.,  et al.\ 2014, A\&A, 562, 67

\reference{guo13}
Guo, Y., Ferguson, H. C., Giavalisco, M., et al.\ 2013, \apjs, 207, 24

\reference{hr93}
Haehnelt, M. G., \& Rees, M. J.\ 1993, \mnras, 263, 168

\reference{hsu14}
Hsu, L.-T., Salvato, M., Nandra, K., et al.\ 2014, \apj, 796, 60 (H14)

\reference{labbe15}
Labb{\'e}, I., Oesch, P. A., Illingworth, G. D., et al.\ 2015, \apjs, 221, 23 

\reference{lodato06}
Lodato, G., \& Natarajan, P.\ 2006, \mnras, 371, 1813

\reference{luo17}
Luo, B., Brandt, W. N., Xue, Y. Q., et al.\ 2017, \apjs, 228, 2 (L17)

\reference{madaurees01}
Madau, P., \& Rees, M. J.\ 2001, \apj, 551, L27

\reference{march16}
Marchesi, S., Civano, M., Salvato, M., et al.\ 2016, \apj, 827, 150

\reference{miller13}
Miller, N. A., Bonzini, M., Fomalont, E. B., et al.\ 2013, ApJS, 205, 13

\reference{nanni17}
Nanni, R., Vignali, C., Gilli, R., Moretti, A., \& Brandt, W. N.\ 2017, A\&A, 603, 128

\reference{pacucci16}
Pacucci, F., Ferrara, A., Grazian, A., et al.\ 2016, \mnras, 459, 1432

\reference{parsa18}
Parsa, S., Dunlop, J. S., \& Mclure, R. J.\ 2018, \mnras, 474, 2904

\reference{planck18}
Planck Collaboration VI, Planck 2018 results. VI. Cosmological parameters. 
2018, A\&A, submitted, arXiv:1807.06209

\reference{poll07}
Polletta, M., Tajer, M., Mareschi, L., et al.\ 2007, \apj, 663, 81

\reference{pons19}
Pons, E., McMahon, R. G., Banerji, M., \& Reed, S. L.\ 2019, \mnras, in press

\reference{rafferty11}
Rafferty, D. A., Brandt, W. N., Alexander, D. M., et al.\ 2011, \apj, 742, 3

\reference{ric19}
Ricarte, A.,  Pacucci, F., Cappelluti, N., et al.\ 2019, \mnras, 489, 1006

\reference{salv19}
Salvestrini, F., Risaliti, G., Bisogni, S., et al.\ 2019, A\&A, 631, 120

\reference{santini15}
Santini, P., Ferguson, H. C., Fontana, A., et al.\ 2015, \apj, 801, 97 (CANDELS)

\reference{santini09}
Santini, P., Fontana, A., Grazian, A., et al.\ 2009, A\&A, 504, 751

\reference{scott10}
Scott, K. S., Yun, M. S., Wilson, G. W., et al.\ 2010, \mnras, 405, 2260

\reference{sijacki15}
Sijacki, D., Vogelsberger, M., Genel, S., et al.\ 2015, \mnras, 452, 575

\reference{skelton14}
Skelton, R. E., Whitaker, K. E., Momcheva, I. G., et al.\ 2014, \apjs, 214, 24

\reference{songaila10}
Songaila, A., \& Cowie, L. L.\ 2010, \apj, 721, 1448

\reference{straatman16}
Straatman, C. M. S., Spitler, L. R., Quadri, R. F., et al.\ 2016, \apj, 830, 51 (ZFOURGE)

\reference{vito16}
Vito, F., Gilli, R., Vignali, C., et al.\ 2016, MNRAS, 463, 348

\reference{vito18}
Vito, F., Brandt, W. N., Yang, G., et al.\ 2018, \mnras, 474, 4528

\reference{vito19a}
Vito, F., Brandt, W. N., Bauer, F., et al.\ 2019a, A\&A, 630, 118

\reference{vito19b}
Vito, F., Brandt, W. N., Yang, G., et al.\ 2019b, A\&A, 629, 6

\reference{volonteri16}
Volonteri, M., Dubois, Y., Pichon, C., \& Devriendt, J.\ 2016, \mnras, 460, 2979

\reference{volonteri03}
Volonteri, M., Haardt, F., \& Madau, P.\ 2003, \apj, 582, 559

\reference{weigel15}
Weigel, A. K., Schawinski, K., Treister, E., et al.\ 2015, \mnras, 448, 3167

\reference{willott11}
Willott, C.\ 2011, \apj, 742, L8

\reference{wyithe11}
Wyithe, J. S. B., \&  Bolton, J. S.\ 2011, \mnras, 412, 1926

\end{references}
\end{document}